\documentclass[12pt,epsfig]{article}
\usepackage[dvips]{graphicx}
\usepackage{amsfonts}
\usepackage{amsmath}
\usepackage{amssymb}
\usepackage[T1]{fontenc}
\usepackage[latin1]{inputenc}
\usepackage{subfigure}
\usepackage{graphicx}

\newtheorem{theorem}{Theorem}[section]
\newtheorem{remark}[theorem]{Remark}

\newtheorem{corollary}[theorem]{Corollary}
\newtheorem{lemma}[theorem]{Lemma}
\newtheorem{proposition}[theorem]{Proposition}
\newtheorem{definition}[theorem]{Definition}

\renewcommand{\theequation} {\thesection.\arabic{equation}}

\begin{document}

\title{Lower Spectral Branches of a Particle Coupled to a Bose
Field}

\author{Nicolae Angelescu $^{1}$, Robert A. Minlos $^{2}$,\\and
\\Valentin A. Zagrebnov $^{3}$}

\date{}

\maketitle

\begin{flushleft}
{\small${}^{1}$  National Institute of Physics and Nuclear Engineering "H. Hulubei",
P.O.Box MG-6, Bucharest, Romania, e-mail: nangel@theory.nipne.ro \\ ${}^{2}$ Institute
for Information Transmissions Problems, Bolshoj Karetny per.19, GSP-4, Moscow 101447,
Russia, e-mail: minl@ittp.ru\\ ${}^{3}$  Universit\'{e} de la M\'editerran\'ee \ and \
Centre de Physique Th\'{e}orique - Luminy, Case 907, Marseille 13288, Cedex 9, France,
e-mail: zagrebnov@cpt.univ-mrs.fr}
\end{flushleft}




\vspace{2em}

{\small\noindent\textbf{Abstract} The structure of the lower part (i.e. $\varepsilon
$-away below the two-boson threshold) spectrum of Fr\"ohlich's polaron Hamiltonian in
the weak coupling regime is obtained in spatial dimension $d\geq 3$. It contains a
single polaron branch defined for total momentum $p\in G^{\left( 0\right) } $, where
$G^{\left( 0\right) }\subset {\mathbb R}^d$ is a bounded domain, and, for any $p\in
{\mathbb R}^d$, a manifold of polaron + one-boson states with boson momentum $q$ in a
bounded domain depending on $p$. The polaron becomes unstable and dissolves into the
one boson manifold at the boundary of $G^{\left( 0\right) }$. The dispersion laws and
generalized eigenfunctions are calculated.}

\section{Introduction}\label{sec:1}

We consider the quantum system consisting of a particle coupled with a Bose field by
an interaction linear in the creation-annihilation operators, known in the physics
literature as Fr\"ohlich's polaron model \cite {Frohlich}. There are many papers, both
physical and mathematical, devoted to this subject, see
\cite{Minlos}-\cite{J.Frohlich}. These are mainly concerned with the ground state
$F_p^{(0)}$ of the Hamiltonian $H_p$ of the system at fixed total momentum $p$ acting
in the Hilbert space ${\cal H}(p)$ (see below). It is shown that for sufficiently
small particle-field coupling constant $\alpha $ the ground state $F_p^{(0)}$ exists
only for momentum $p$ in a certain domain $G^{\left( 0\right) }\subset {\mathbb R}^d$,
where $G^{\left( 0\right) }$ is bounded for space dimension $d\geq 3$ and $G^{\left(
0\right) }={\mathbb R}^d$ for $d=1,2$. The ground state describes the ''polaron'',
i.e. the particle in a ''cloud of virtual bosons''.

Here we study the next, ''one-boson'', branch of the spectrum of $H_p$ for $%
d\geq 3$. The expected mathematical picture is the following: there exists an invariant
subspace ${\cal H}_1(p)\subset {\cal H}(p)$ of the operator $H_p$, which is
isomorphic in a natural way with $L_2\left( G_p^{(1)},dq\right) $, where $%
G_p^{(1)}\subset {\mathbb R}^d$ is a certain bounded domain, such that $H_p$ acts in
this subspace as multiplication with a function $\xi _p\left( q\right) $, which can be
viewed as the energy of a boson of momentum $q$ (while the total momentum of the
system is $p$). The range of this function is the segment $\left[ \lambda _1\left(
p\right) ,\lambda _2\left( p\right) \right) $, where $\lambda _1\left( p\right) $ and
$\lambda _2\left( p\right) $ are the thresholds of the one- and two-boson states,
respectively. Moreover, in the subspace orthogonal to ${\cal H}_0(p)\oplus {\cal
H}_1(p)$, where ${\cal H}_0(p)=\left\{ cF_p^{(0)}\right\} $ is the one-dimensional
subspace generated by the ground state whenever it exists, the spectrum of $%
H_p$ lies above $\lambda _2\left( p\right) $ (this latter property will be
called ''the completeness of the one-boson spectrum''). The states in ${\cal %
H}_1(p)$ can be viewed as scattering states of a boson and a polaron.

Unfortunately, we shall obtain here only part of the above picture. Namely,
we are able to construct only a subspace ${\cal H}_1^\kappa (p)\subset {\cal %
H}_1(p)$ isomorphic to $L^2\left( G_p^{(1),\kappa },dq\right) $, where $%
G_p^{(1),\kappa }=\left\{ q\in G_p^{(1)}:\xi _p(q)<\kappa \right\} $. Here, $%
\kappa <\lambda _2\left( p\right) $ can be chosen arbitrarily close to $%
\lambda _2\left( p\right) $, at the expense of taking the coupling constant $%
\alpha $ sufficiently small. Apparently, our techniques allow the
construction of the whole space $G_p^{(1)}$ and the proof of the
completeness of the one-boson spectrum for sufficiently large space
dimension $d$.

Our analysis of the one-boson branch covers only the cases $d\geq 3$, though we expect
that the same picture holds in lower dimension, with $G_p^{\left( 1\right) }={\mathbb
R}^d$ for $d=1,2$. The calculations are based on a technique used by one of the
authors in \cite{Minlos}, and also on certain facts connected with the spectral
analysis of the so-called generalized Friedrichs model \cite{friedrichs}.

We proceed now to a detailed presentation of the model and a precise statement of the
main result.

The state space of our model is the Hilbert space $$ {\cal H}=L^2({\mathbb R}^d)
\otimes {\cal F}, $$ where ${\cal F}$ is the symmetric (boson) Fock space $$ {\cal
F}={\cal F}_{{sym}}\left( L^2({\mathbb R}^d) \right) =\bigoplus\limits_{n=0}^\infty
{\cal H}^{\left( n\right) }, $$ with ${\cal H}^{\left( 0\right) }={\mathbb C}$, ${\cal
H}^{\left( n\right) }=\left( L^2({\mathbb R}^d) \right) _{{sym}}^{\otimes n}$ the
symmetric tensor power ($n\geq 1$). Thus, the vectors of ${\cal H}$ are sequences
\begin{equation}
\label{eq. 1.1}F=\left\{ f_0\left( p_0\right) ,f_1\left( p_0;q\right)
,...,f_n\left( p_0;q_1,...,q_n\right) ,...\right\} ,
\end{equation}
where $f_n$ are, for every $p_0\in {\mathbb R}^d$, symmetric functions of the
variables $q_1,...,q_n$, and the norm is given by
\begin{equation}
\label{eq. 1.2}\left\| F\right\| ^2=\int_{{\mathbb R}^d}\left| f_0\left(
p_0\right) \right| ^2dp_0+\sum\limits_{n=1}^\infty \frac 1{n!}\int_{{\mathbb R}%
^d}\int_{{\mathbb R}^{nd}}\left| f_n\left( p_0;q_1,...,q_n\right) \right|
^2dp_0\prod\limits_{i=1}^ndq_i.
\end{equation}
The Hamiltonian of our system has the form
\begin{equation}
\label{eq. 1.3}H=H_0^{{part}}+H_0^{{field}}+\alpha H_{{int}},
\end{equation}
where $\alpha >0$ is a coupling constant and
\begin{equation}
\label{eq. 1.4}
\begin{array}{l}
\left( H_0^{
{part}}F\right) _n\left( p_0;q_1,...q_n\right) =\frac 12p_0^2f_n\left(
p_0;q_1,...,q_n\right) , \\  \\
\left( H_0^{
{field}}F\right) _n\left( p_0;q_1,...q_n\right) =\left(
\sum\limits_{i=1}^n\varepsilon \left( q_i\right) \right) f_n\left(
p_0;q_1,...,q_n\right) , \\  \\
\left( H_{
{int}}F\right) _n\left( p_0;q_1,...q_n\right)
=\sum\limits_{i=1}^nc(p_0;q_i)f_{n-1}\left( p_0+q_i;q_1,...,\check
q_i,...,q_n\right) \\ \hfill+\int_{{\mathbb R}^d}\overline{c(p_0;q)}%
f_{n+1}\left( p_0-q;q_1,...q_n,q\right) dq  ,
\end{array}
\end{equation}
with the convention that a sum over a void set is 0, and where the notation $\check q$
means that the variable $q$ is omitted. The properties of the functions $\varepsilon $
and $c$ will be given in detail later. Notice that, with the minimal assumptions:
$\varepsilon $ is a positive real function and the function $c$ is bounded and with
sufficiently rapid decay for $q\rightarrow \infty $, the operator $H$ is self-adjoint
and bounded from below.

A first simplification in the spectral analysis of $H$ comes from the
conservation of the total momentum, i.e. from the fact that $H$ commutes
with the operator
\begin{equation}
\label{eq. 1.5}\left( \hat PF\right) _n\left( p_0;q_1,...q_n\right) =\left(
p_0+\sum\limits_{i=1}^nq_i\right) f_n\left( p_0;q_1,...,q_n\right) ,n\geq 0.
\end{equation}
As a consequence, ${\cal H}$ can be written as a direct integral of Hilbert
spaces ${\cal H}\left( p\right) $%
\begin{equation}
\label{eq. 1.6}{\cal H}=\int_{{\mathbb R}^d}^{\oplus }{\cal H}\left( p\right) dp,
\end{equation}
which reduces both $\hat P$ and $H$, i.e. induces the decompositions
\begin{equation}
\label{eq. 1.7}\hat P=\int_{{\mathbb R}^d}^{\oplus }pIdp,\;\;H=\int_{{\mathbb R}%
^d}^{\oplus }H_pdp,
\end{equation}
where $I$ (the unit operator) and $H_p$ are operators in ${\cal H}\left( p\right) $. For
a vector $F$ as given in (\ref{eq. 1.1}), we get the representation: $$ F=\int_{{\bf
R}^d}^{\oplus }\hat F_pdp, $$ where $\hat F_p=\left\{ \hat f_{p,n}\right\}_{n\geq 0} $
and $\hat f_{p,n}$ is the restriction of $f_n$ to the hyperplane
$p_0+\sum\limits_{i=1}^nq_i=p$.
The spaces ${\cal H}\left( p\right) $ will be identified with ${\cal F}=%
{\cal F}_{{sym}}\left( L^2({\mathbb R}^d) \right) $ by means of the unitaries
\begin{equation}
\label{eq. 1.8}\left( U_p\hat F_p\right) _n\left( q_1,...q_n\right)
=f_n\left( p-\sum\limits_{i=1}^nq_i;q_1,...,q_n\right) .
\end{equation}
With this identification, the action of $H_p$ in ${\cal F}$ is given
by the formula
\begin{equation}
\label{eq. 1.9}
\begin{array}[b]{ll}
\left( H_pF\right) _n\left( q_1,...q_n\right) & =e_{n,p}^0\left(
q_1,...,q_n\right) f_n\left( q_1,...,q_n\right) \\
& +\alpha \sum\limits_{i=1}^nc(p-\sum\limits_{j=1}^nq_j;q_i)f_{n-1}\left(
q_1,...,\check q_i,...,q_n\right) \\
& +\alpha \int\limits_{{\mathbb R}^d}\overline{c(p-q-\sum\limits_{j=1}^nq_j;q)}%
f_{n+1}\left( q_1,...q_n,q\right) dq,
\end{array}
\end{equation}
where
\begin{equation}
\label{eq. 1.10}e_{n,p}^0\left( q_1,...,q_n\right) =\frac 12\left(
p-\sum\limits_{i=1}^nq_i\right) ^2+\sum\limits_{i=1}^n\varepsilon \left(
q_i\right) ,
\end{equation}

The functions $\varepsilon $ and $c$ are supposed to fulfill the following
conditions:
\begin{enumerate}
\item  $\varepsilon (q)$ {\em is a convex, non-decreasing function of }$%
\left| q\right| ${\em \ and there exists }$c_o>0${\em , such that}
\begin{equation}
\label{eq. 1.11}\varepsilon (q_1)+\varepsilon (q_2)\geq \varepsilon
(q_1+q_2)+c_o,\;\;\forall q_1,q_2\in {\mathbb R}^d.
\end{equation}
Also, we need stronger regularity properties: $\varepsilon \in C^\infty \left(
{\mathbb R}^d\right) ${\em \ and it has bounded derivatives, i.e. there exists
}$R>0${\em , such that for all multi-indices} $\alpha =\left( \alpha _1,...,\alpha
_d\right) \neq 0$,
\begin{equation}
\label{eq. 1.12}\sup _{q\in {\mathbb R}^d}\left| \partial _q^\alpha \varepsilon \left(
q\right) \right| \leq R,
\end{equation}
{\em where}%
$$
\partial _q^\alpha =\frac{\partial ^{\left| \alpha \right| }}{\partial
q_1^{\alpha _1}...\partial q_d^{\alpha _d}},\;q=(q_1,...,q_d),\;\;\left|
\alpha \right| =\sum\limits_{i=1}^d\alpha _i.
$$
The following are physically interesting examples of such functions:%
$$
\begin{array}{ll}
a)\;\varepsilon \left( q\right) =\varepsilon \left( 0\right) >0 &  \\
b)\;\varepsilon \left( q\right) =\sqrt{q^2+m^2}+c_o, & m>0 ,\, c_o
>0
\end{array}
$$
\item  $c(p,q)$ {\em is sufficiently smooth and there exists a bounded,
rapidly decreasing function }$h:{\mathbb R}^d\rightarrow {\mathbb R}_{+}${\em \
dominating }$c${\em \ and all its derivatives, i.e., for all multi-indices }$%
\alpha ,\beta ${\em \ there exists }$C_{\alpha ,\beta }>0${\em \ ($C_{0,0}=1$%
), such that}
\begin{equation}
\label{eq. 1.13}\left| \partial _p^\alpha \partial _q^\beta c\left( p;q\right) \right|
\leq C_{\alpha ,\beta }h(q),\;\forall p,q\in {\mathbb R}^d.
\end{equation}
\end{enumerate}

We are concerned here with the study by perturbation theory of the
(lower part of the)
spectrum of the Hamiltonian (\ref{eq. 1.9}) for every fixed $p$:
\begin{equation}
\label{eq. 1.14}H_p=H_p^{(0)}+\alpha H_{p,{int}},
\end{equation}
where $H_p^{(0)}$ denotes the first term, and $\alpha H_{p,{int}}$ the other terms, of
equation(\ref{eq. 1.9}); sometimes, for notational simplicity, the index $p$ will be
omitted.

The spectrum of $H_p^{\left( 0\right) }$ consists of the eigenvalue
$\frac 12p^2$, corresponding to the bare particle, and branches of
continuous spectrum $e_{n,p}^0\left( q_1,...,q_n\right) $,
corresponding to bare particle + $n$-boson states, starting at the
thresholds
\begin{equation}
\label{eq. 1.15}\lambda _n^0(p)=\min _{q_1,...,q_n}e_{n,p}^0\left(
q_1,...,q_n\right) .
\end{equation}
Remark that, in view of the convexity of $p^2$ and $\varepsilon $,
the minimum of $e_{1,p}^0\left( q_1\right) $ is attained at a single
point $q_1^0 $, which is its unique critical point and is
nondegenerate. Moreover, as a consequence of the inequality
(\ref{eq. 1.11}),
\begin{equation}
\label{eq. 1.16}\lambda _n^0(p)\geq \lambda _{n-1}^0(p)+c_o.
\end{equation}

The main result of the paper is summarized in the following:
\begin{theorem}\label{T1.1} \hfill
\newline
1) For any $d\geq 3$, there exists $\alpha _0=\alpha _0(d)$
\ such that, for any $\alpha <\alpha _0$ \ there exist functions $%
\lambda _1(p)<\lambda _2(p)$ , with $\lambda _1(p)<\lambda _1^0(p),\,  \lambda
_2(p)<\lambda _2^0(p)$  , and a bounded domain $G^{(0)}\subset {\mathbb R}^d$ , such
that the spectrum of $H_p$ \ in $\left( -\infty ,\lambda _1(p)\right) $ \ consists of
one nondegenerate eigenvalue $\xi _p^{\left( 0\right)} $
if $p\in G^{(0)}$ \ and is void if $p\notin G^{(0)}$ . Moreover, $%
\xi _p^{\left( 0\right) }<p^2/2$ \ and $\lim \limits_{p^{\prime
}\rightarrow p\in \partial G^{(0)}}\xi _{p^{\prime }}^{\left(
0\right) }=\lambda _1(p)$ , where $\partial G^{(0)}$ \ denotes the
boundary of the domain $G^{(0)}$ . The associated eigenvector
$F_p^{(0)}$ \ is the ground state of $H_p$ .\\
\noindent 2) For any $\kappa \in \left( \lambda _1(p),\lambda _2(p)\right) $ %
\ and any $p\in {\mathbb R}^d$ , there exists $\bar \alpha _0=\bar \alpha _0\left(
\kappa ,p,d\right) $, such that for any $\alpha
<\bar \alpha _0$ \ there exists a domain $G_p^{(1),\kappa }\subset
{\mathbb R}^d$ , a $C^\infty $-function $\xi _p^\kappa :G_p^{(1),\kappa}
\rightarrow \left[\lambda _1 (p),\kappa \right]$  and a subspace ${\cal H}%
_1^\kappa (p)\subset {\cal F}$ \ invariant for $H_p$ , such that the
restriction of $H_p$ \ to ${\cal H}_1^\kappa (p)$ \ is unitarily
equivalent to the operator of multiplication by the function $\xi
_p^\kappa $\ in $L^2\left( G_p^{(1),\kappa },dq\right) $ . Thereby,
for $\kappa _1<\kappa _2\in \left( \lambda _1(p),\lambda _2(p)\right) $  one gets \ $%
G_p^{(1),\kappa _1}\subset G_p^{(1),\kappa _2}$ \ and \ $\xi _p^{\kappa _1}=\xi
_p^{\kappa _2}\mid _{G_p^{(1),\kappa _1}}$ .
\end{theorem} \vspace{0.2cm}
\begin{remark}\label{R1.1 } Refining slightly the technique of this paper, one can reach $%
\kappa =\lambda _2(p)$ if the dimension $d$ is sufficiently large, i.e. $%
\bar \alpha _0\left( \cdot ,p,d\right) $ is bounded away from zero,
and the whole one-boson subspace ${\cal H}_1^{\kappa =\lambda
_2(p)}(p)$ and the function  $\xi _p^{\kappa =\lambda _2(p)}$ can be
constructed.
\end{remark}

\section{Outline of the proof\label{outline}}

\setcounter{equation}{0}

We shall present first the strategy we adopt in proving Theorem \ref{T1.1}, in order not
to obscure it by the cumbersome calculations to be done.

Our constructions involve the resolvent of $H_p:$%
\begin{equation}
\label{eq. 1.17}R(z)=\left( H_p-zI\right) ^{-1}.
\end{equation}
We have therefore to solve, for any $L\in {\cal F}$, the equation:
\begin{equation}
\label{eq. 1.18}\left( H_p-zI\right) F=L.
\end{equation}
We split the space ${\cal F}$ as an orthogonal sum ${\cal H}^{(\leq
1)}\oplus {\cal H}^{(\geq 2)}$, corresponding to number $n$ of bare Bosons $%
\leq 1$, and $\geq 2$, respectively, and denote $\Pi _1,\Pi _2$ the
corresponding orthogonal projections. Hence, $F=F_1+F_2$, where $F_1=\Pi _1F$%
=$\left\{ f_0,f_1,0,0,...\right\} $, $F_2=\Pi _2F=\left\{ 0,0,f_2,...\right\} $ and
similarly the vector $L=L_1+L_2$. Then the operator $H_p$ has a matrix representation:
\begin{equation}
\label{eq. 1.19}H_p=\left(
\begin{array}{cc}
A_{11} & \alpha A_{12} \\ \alpha A_{21} & A_{22}
\end{array}
\right) ,
\end{equation}
where $A_{ii}=\Pi _iH_p\Pi _i$, $\alpha A_{ij}=\Pi _iH_p\Pi _j$
($i\neq j$), in terms of which equation (\ref{eq. 1.18}) writes as
\begin{equation}
\label{eq. 1.20}\left\{
\begin{array}{c}
\left( A_{11}-zI\right) F_1+\alpha A_{12}F_2=L_1 \\ \alpha A_{21}F_1+\left(
A_{22}-zI\right) F_2=L_2
\end{array}
\right. .
\end{equation}

We define
\begin{equation}
\label{eq. 1.21}\lambda _2(p)=\inf \;spec\left( A_{22}\right) .
\end{equation}
By the variational principle,
\begin{equation}
\label{eq. 1.22}\lambda _2(p)=\inf \limits_{F\in {\cal F};\left\| F\right\| =1}\left(
\Pi _2F,H_p\Pi _2F\right) \leq \lambda _2^0(p)
\end{equation}
as $\lambda _2^0(p)$ is obtained as the infimum of the same
expression taken over the subspace ${\cal H}^{(\leq 2)}$ of vectors
$F$ with at most two bare bosons. Likewise, we define
\begin{equation}
\label{eq. 1.22'}\lambda _1(p)=\inf \limits_{F\in {\cal H}^{(\geq
1)};\left\| F\right\| =1}\left( F,H_pF\right) \leq \lambda _1^0(p)
\end{equation}
the infimum of the spectrum of the restriction of $H_p$ to the subspace $%
{\cal H}^{(\geq 1)}$ with at least one bare boson.

For $z$ in the resolvent set of $A_{22}$, the second equation in (\ref
{eq.
1.20}) can be solved for $F_2$%
\begin{equation}
\label{eq. 1.23}F_2=\left( A_{22}-zI\right) ^{-1}\left( L_2-\alpha
A_{21}F_1\right) ,
\end{equation}
and hence one arrives at the following equation for $F_1$%
\begin{equation}
\label{eq. 1.24}\left( A_{11}-\alpha ^2A_{12}\left( A_{22}-zI\right)
^{-1}A_{21}-zI\right) F_1=L_1-\alpha A_{12}\left( A_{22}-zI\right) ^{-1}L_2.
\end{equation}

Let now restrict to real $z =\xi$ and consider the family of
self-adjoint operators acting in ${\cal H}^{(\leq 1)}$:
\begin{equation}
\label{eq. 1.25}A_p\left( \xi \right) =A_{11}-\alpha ^2A_{12}\left(
A_{22}-\xi I\right) ^{-1}A_{21},\;\xi \in \left( -\infty ,\lambda
_2(p)\right) .
\end{equation}

We shall show that, under our assumptions and for $\xi \leq \kappa \in \left( \lambda
_1(p),\lambda _2(p)\right) $, $A_p\left( \xi \right) $ are  generalized Friedrichs
operators, i.e. each operator $A_p\left( \xi \right) =A $ allows in the space ${\cal
H}^{(\leq 1)}={\mathbb C}\oplus L^2({\mathbb R}^d,dq)$ a representation of the form:

\begin{equation}
\label{eq. 1.26}
\begin{array}{l}
\left( AF\right) _0=e^{\left( 0\right) }f_0+\alpha \int \bar v\left(
q\right) f_1\left( q\right) dq \\
\left( AF\right) _1=\alpha v\left( q\right) f_0+a\left( q\right) f_1\left(
q\right) +\alpha ^2\int D\left( q,q^{\prime }\right) f_1\left( q^{\prime
}\right) dq^{\prime },
\end{array}
\end{equation}
where $F=\left( f_0,f_1\right) \in {\cal H}^{(\leq 1)}$. Here $v\left( q\right) $,
$a\left( q\right) $ and the kernel $D\left( q,q^{\prime }\right) $ fulfill a set of
smoothness conditions (given in detail in Section \ref{sec:fried}), $a\left( q\right) $
is bounded from below and grows at most linearly at $\infty $, and $v\left( q\right) $,
$D\left( q,q^{\prime }\right) $ decrease fast at $\infty $. Such operators allow, for
small $\alpha $, a complete spectral analysis (see \cite{friedrichs}-\cite{Reed} and
Section \ref{sec:fried} below), namely, letting aside the possible eigenvalue $e_p$
(ground state), they are unitarily
equivalent to the operator of multiplication by $a\left( q\right) $ in $%
L_2\left( {\mathbb R}^d,dq\right) $.

Let us denote, for given $p$ and $\xi $, by $a_p\left( \xi ,q\right) $ the
function $a\left( q\right) $ entering equation (\ref{eq. 1.26}) written for $%
A_p\left( \xi \right) $. In essence, the key to the spectral analysis of $%
H_p $ lies the following remark:
\begin{remark}\label{R2.1}  Let $F=F_1 + F_2$ \ ($F_1\in {\cal
H}^{(\leq
1)},F_2\in {\cal H}^{(\geq 2)}$) be (generalized) eigenvector of $H_p$%
 \ with eigenvalue $\xi .$  Then, by equation (\ref{eq. 1.24}), $F_1$%
 \ is a (generalized) eigenvector of the operator $A_p\left( \xi
\right) $\ with the same eigenvalue $\xi $ . Conversely, suppose
that $F_{\xi ,1}$ \ is the eigenvector of $A_p\left( \xi \right) $\
of eigenvalue $e_p\left( \xi \right) $ \ (whenever it exists). If
the equation
\begin{equation}
\label{eq. 1.27}e_p\left( \xi \right) =\xi
\end{equation}
 has a solution $\xi _p^{\left( 0\right) }$, then $F=F_{\xi
_p^{\left( 0\right) },1} + F_{\xi _p^{\left( 0\right) },2}$ , where
$$
F_{\xi _p^{\left( 0\right) },2}=-\alpha \left( A_{22}-\xi _p^{\left(
0\right) }I\right) ^{-1}A_{21}F_{\xi _p^{\left( 0\right) },1}
$$
is an eigenvector of $H_p$ \ with eigenvalue $\xi _p^{\left(
0\right) }$ . Likewise, let for a given $p$,  $F_{\xi ,1}^q$ %
\ be a generalized eigenvector of the operator $A_p\left( \xi \right) $\ corresponding
to the eigenvalue $a_p\left( \xi ,q\right) $ \ and $\xi _p\left( q\right) $ \ be a
solution of the equation
\begin{equation}
\label{eq. 1.27'}a_p\left( \xi ,q\right) =\xi .
\end{equation}
Then for $F_1^q=F_{\xi \left( q\right) ,1}^q $  \ and
\begin{equation}
\label{eq. 1.50}F_2^q=-\alpha \left( A_{22}-\xi \left( q\right)
I\right) ^{-1}A_{21}F_1^q \,,
\end{equation}
\textit{the vector} $F^q=F_1^q+F_2^q$  is a generalized eigenvector
of $H_p$ \ for the eigenvalue $\xi _p\left( q\right) $ .
\end{remark}

The domain $G^{(0)}$ is identified with the set of $p$ for which
equation (\ref {eq. 1.27}) has a solution. For any given $p$,
$G_p^{(1),\kappa }$ is the set of $q$ for which equation (\ref{eq.
1.27'}) has a solution $\xi \left( q\right) \leq \kappa $. The
constructions of the subspace ${\cal H}_1^\kappa (p)$ and of the
unitary equivalence of the operator $H_p\mid _{{\cal H}_1^\kappa
(p)}$ to the multiplication by $\xi _p^\kappa (q)=\xi _p(q)$ are
done in the standard way in terms of the family $\left\{
F^q\right\}_{ q\in G_p^{(1),\kappa }} $ of generalized eigenvectors
of $H_p$.

\section{Elimination of the many-body components}

\setcounter{equation}{0}

In this section we study perturbatively the solution (\ref{eq.
1.23}) and derive its main properties of interest to us. By
virtue of equation (\ref{eq. 1.14}) and the inequality (\ref{eq.
1.22}), one can factorize the unperturbed (diagonal) part
$H_p^{\left( 0\right) }-z$ for $z\notin \left[ \lambda _2(p),\infty
\right) $, and bring the second equation (\ref{eq. 1.20}) to the
form of the equivalent fixed-point equation:
\begin{equation}
\label{eq. 2.1}F_2+Q(z)F_2=\alpha \left( H_p^{\left( 0\right)
}-z\right) ^{-1}L_2-GF_1,
\end{equation}
where
\begin{equation}
\label{eq. 2.2}Q(z)=\alpha \left( H_p^{\left( 0\right) }-z\right) ^{-1}\Pi
_2H_{p,int}\Pi _2,
\end{equation}
\begin{equation}
\label{eq. 2.3}G=\alpha \left( H_p^{\left( 0\right) }-z\right)
^{-1}\Pi _2H_{p,int}\Pi _1 \,,
\end{equation}
 Explicitly, the vector
$GF_1$ has the form $\left\{ 0,0,g_2,0,...\right\} $ with
\begin{equation}
\label{eq. 2.4}g_2(q_1,q_2)=\frac{\alpha \left[ c\left( p-q_1-q_2;q_1\right)
f_1(q_2)+c\left( p-q_1-q_2;q_2\right) f_1(q_1)\right] }{e_{2,p}^0\left(
q_1,q_2\right) -z}.
\end{equation}
\begin{lemma}\label{L3.1} For every $\kappa \in \left(
\lambda _1(p),\lambda _2(p)\right)$ there exists $\alpha _0\left(
\kappa \right) $ such that, for any $\alpha <\alpha _0\left( \kappa
\right) $, and any  $z\in {\mathbb C}$ with $\mathfrak{Re}\,z=\xi
\leq \kappa $,
 $\left\| Q(z)\right\| <1/2$, therefore equation (\ref{eq. 2.1})
has a unique solution $F_2$ for every $f_1\in L^2({\mathbb R}^d,dq) $ and $ L_2\in
{\cal H}^{(\geq 2)}$ .
\end{lemma}
\textit{Proof} : We write $Q\left( z\right) $ as a sum of its
creation and annihilation parts: $Q\left( z\right) =Q^{\prime
}+Q^{\prime \prime }$, with
\begin{equation}
\label{eq. 2.5}
\begin{array}{l}
\left( Q^{\prime }F\right) _n\left( q_1,...,q_n\right) \\
=\left\{
\begin{array}{ll}
0, & n=2 \\
\alpha \left( e_{n,p}^0\left( q_1,...,q_n\right) -z\right)
^{-1}\sum\limits_{i=1}^nc\left( p-\sum_jq_j;q_i\right) f_{n-1}(...\check
q_i...), & n>2
\end{array}
\right.
\end{array}
\end{equation}
\begin{equation}
\label{eq. 2.6}
\begin{array}{l}
\left( Q^{\prime \prime }F\right) _n\left( q_1,...,q_n\right) \\
=\alpha \left( e_{n,p}^0\left( q_1,...,q_n\right) -z\right) ^{-1}\int
\overline{c\left( p-\sum_jq_j-q;q\right) }f_{n+1}(q_1,...,q_n,q)dq, \\
\hfill n\geq 2
\end{array}
\end{equation}
By condition (\ref{eq. 1.11}) and the inequality (\ref{eq. 1.22}),
one gets for $\xi
\leq \kappa $%
\begin{equation}
\label{eq. 2.7}\left| e_{n,p}^0\left( q_1,...,q_n\right) -z\right| \geq
e_{n,p}^0\left( q_1,...,q_n\right) -\xi \geq \left( n-2\right) c_o+\lambda
_2\left( p\right) -\kappa ,
\end{equation}
therefore, by virtue of (\ref{eq. 1.13}),
$$
\left\| \left( Q^{\prime }F\right) _n\right\| _{L^2\left( {\mathbb R}%
^{dn}\right) }\leq \frac{n\alpha \cdot \left\| h\right\| _{L^2\left( {\mathbb R}%
^d\right) }\cdot \left\| f_{n-1}\right\| _{L^2\left( {\bf
R}^{d(n-1)}\right) }}{\left( n-2\right) c_o+\lambda _2\left(
p\right) -\kappa },\;n>2,
$$
while, for $n=2$, the norm vanishes. Therefore,
$$
\begin{array}{l}
\left\| Q^{\prime }F\right\| _{ {\cal H}^{(\geq 2)}}^2=\sum\limits_{n=2}^\infty \frac
1{n!}\cdot \left\| \left( Q^{\prime }F\right) _n\right\| _{L^2\left( {\mathbb
R}^{dn}\right) }^2 \\ \leq \alpha ^2\left\| h\right\| _{L^2({\mathbb R}^d) }^2\max
_{n\geq 3}\left[ n\left( \left( n-2\right) c_o+\lambda _2\left( p\right) -\kappa
\right) ^{-2}\right] \\ \hfill {\times \sum\limits_{n=3}^\infty }\frac 1{\left(
n-1\right) !}\cdot \left\| f_{n-1}\right\| _{L^2\left( {\mathbb R}^{d(n-1)}\right) }^2
\\ \leq \alpha ^2\left\| h\right\| _{L^2({\mathbb R}^d) }^23\left( c_o+\lambda _2\left(
p\right) -\kappa \right) ^{-2}\left\| F\right\| _{{\cal H}^{(\geq 2)}}^2,
\end{array}
$$ implying that $\left\| Q^{\prime }\right\| \leq \alpha \cdot \left\| h\right\|
_{L^2({\mathbb R}^d) }\sqrt{3}/\left( c_o+\lambda _2\left( p\right) -\kappa \right) $.
A similar calculation shows that $\left\| Q^{\prime \prime }\right\| \leq \alpha \cdot
\left\| h\right\| _{L^2({\mathbb R}^d) }\sqrt{3}/\left( \lambda _2\left( p\right)
-\kappa \right) $. This finishes the proof of the lemma. \hfill $\square$

Let us denote by $F_2^0(z,f_1)=\left\{ f_n^0\left( z;\cdot \right)
;n\geq 2\right\} $ the solution of equation (\ref{eq. 2.1}) for
$L_2=0$. Under the conditions of Lemma 3.1 and taking into account equation (\ref{eq.
2.4}), we get that $ \left\| F_2^0(z,f_1)\right\| _{{\cal H}^{(\geq 2)}}\leq C\alpha
\left\| f_1\right\| _{L^2({\mathbb R}^d) }$. From now on, we shall denote by $S(z)$
the linear operator:
\begin{equation}
\label{eq. 2.7a}{\cal H}^{(\leq 1)}\ni f_1\stackrel{S(z)}{
\longmapsto }F_2^0(z,f_1)\in {\cal H}^{(\geq 2)}.
\end{equation}

To proceed further with the analysis we need more information about
the structure and regularity of the solution $F_2^0(z,f_1)$. To this
aim, we shall solve equation (\ref{eq. 2.1}) with $L_2=0$. In
particular, we shall show that the components of $F_2^0(z,f_1)$ have
the representation
\begin{equation}
\label{eq. 2.8}
\begin{array}{r}
f_n^0\left( z;q_1,...,q_n\right) =\sum\limits_{i=1}^nb_n\left(
q_1,...,\check q_i..,q_n;q_i\right) f_1\left( q_i\right) \\
+\int d_n\left( q_1,...,q_n;q\right) f_1\left( q\right) dq,
\end{array}
\end{equation}
where the functions $b_n\left( q_1,...,q_{n-1};q_n\right) $ are
symmetric in $q_1,...,q_{n-1}$ and the functions $d_n\left(
q_1,...,q_n;q\right) $ are symmetric in $q_1,...,q_n$, $n\geq 2$.
The functions $b_n$ and $d_n$ will be called \textit{coefficient
functions}.

A simple calculation shows that, if $F\in {\cal H}^{(\geq 2)}$ has the
representation (\ref{eq. 2.8}), then also $\hat F=Q\left( z\right) F$ has
the same kind of representation, with coefficient functions
\begin{equation}
\label{eq. 2.9b}
\begin{array}{l}
\hat b_n\left( q_1,...,q_{n-1};q_n\right) =\alpha \left( e_{n,p}^0\left(
q_1,...,q_n\right) -z\right) ^{-1} \\
\times \left[ \sum\limits_{i=1}^{n-1}c\left(
p-\sum\limits_{j=1}^nq_j;q_i\right) b_{n-1}\left( q_1,...,\check
q_i..,q_{n-1};q_n\right) \right. \\
\left. +\int \bar c\left( p-\sum\limits_{j=1}^nq_j-q;q\right)
b_{n+1}\left( q_1,...,q_{n-1},q;q_n\right) dq\right]
\end{array}
\end{equation}

\begin{equation}
\label{eq. 2.9d}
\begin{array}{l}
\hat d_n\left( q_1,...,q_n;q\right) =\alpha \left( e_{n,p}^0\left(
q_1,...,q_n\right) -z\right) ^{-1} \\ \times \left[
\sum\limits_{i=1}^nc\left( p-\sum\limits_{j=1}^nq_j;q_i\right)
d_{n-1}\left( q_1,...,\check q_i..,q_n;q\right) \right. \\ +\int
\bar c\left( p-\sum\limits_{j=1}^nq_j-q^{\prime };q^{\prime }\right)
d_{n+1}\left( q_1,...,q_n,q^{\prime };q\right) dq^{\prime } \\ +
\left. \overline{c}\left( p -\sum\limits_{j=1}^nq_j-q;q\right)
b_{n+1}\left( q_1,...,q_n;q\right) \right]
\end{array}
\end{equation}

Let now define the space ${\cal M}$ of all pairs $\mu =\left\{ \left( b_n\right)
_{n\geq 2},\left( d_n\right) _{n\geq 2}\right\}$ of sequences of bounded continuous
functions, $b_n:\left( {\mathbb R}^d\right) ^{\left(
n-1\right) }\times {\mathbb R}^d\rightarrow {\mathbb C}$, $d_n:\left( {\mathbb R}%
^d\right) ^n\times {\mathbb R}^d\rightarrow {\mathbb C}$, symmetric with respect to
the first group of variables. Let them fulfill the following condition: there exists a
constant $M$, such that,
\begin{equation}
\label{eq. 2.10}
\begin{array}[b]{l}
\sup _q\left| b_n\left( q_1,...,q_{n-1};q\right) \right| \leq M
\prod\limits_{i=1}^{n-1}h\left( q_i\right) , \\ \left| d_n\left( q_1,...,q_n;q\right)
\right| \leq M h\left( q\right) \prod\limits_{i=1}^nh\left( q_i\right) ,\hskip 6mm
\forall n\geq 2
\end{array}
\end{equation}
where $h$ is the function appearing in equation (\ref{eq. 1.13}).
${\cal M}$ is a Banach space with the norm
\begin{equation}
\label{eq. 2.11}\left\| \mu \right\| =\inf M,
\end{equation}
where the infimum is taken over all $M$ for which the condition
(\ref {eq. 2.10}) holds. Clearly, equation (\ref{eq. 2.8}) defines a
continuous application of ${\cal H}^{(1)}$ into ${\cal H}^{(\geq
2)}$.

The linear operator $\Gamma \left( z\right) $ acting in ${\cal M}$ according to $\Gamma
\left( z\right) \mu =\hat \mu $, where $\mu =\left\{ \left( b_n\right) _{n\geq 2},\left(
d_n\right) _{n\geq 2}\right\} $ and $\hat \mu =\{(\hat b_n) _{n\geq 2},(\hat d_n)
_{n\geq 2}\} $ are related by (\ref{eq. 2.9b}), (\ref{eq. 2.9d}), translates in ${\cal
M}$ the action of $Q(z)$. Then equation (\ref{eq. 2.1}) with $L_2=0$ is transformed into
\begin{equation}
\label{eq. 2.12}\mu +\Gamma \left( z\right) \mu =\mu _0,
\end{equation}
where $\mu _0=\left\{ \left( b_n^0\right) ,\left( d_n^0\right) \right\} $
with $b_n^0=0$, $\forall n\geq 3$, $d_n^0=0$, $\forall n\geq 2$, and
\begin{equation}
\label{eq. 2.13}b_2^0\left( q_1;q\right) = -\frac{\alpha c\left( p-q_1-q;q_1\right)
}{e_{2,p}^0\left( q_1,q\right) -z}.
\end{equation}
\begin{lemma}\label{L3.2} For every $\kappa \in \left( \lambda
_1(p),\lambda _2(p)\right) $ there exists $\tilde \alpha _0\left(
\kappa \right) $ such that, for any $\alpha <\tilde \alpha _0\left(
\kappa \right) $, and any $z\in {\mathbb C}$ with $\mathfrak{Re}
\,z=\xi \leq \kappa $, $ \left\| \Gamma (z)\right\| <1/2$, and $\mu
_0\in {\cal M}$, $ \left\| \mu _0\right\| \leq \alpha /\left(
\lambda _2(p)-\kappa \right) $. Therefore, equation (\ref{eq. 2.12})
has a unique solution $\mu \left( z\right) \in {\cal M}$, which is
an analytic function of $z$ in the half-plane $\mathfrak{Re}\,z\leq
\kappa $. Moreover, for any $r\geq 1$,
there exists $\tilde \alpha _r\left( \kappa \right) $, such that, for $%
\alpha <\tilde \alpha _r\left( \kappa \right) $, the components of
$\mu \left( z\right) $ are $C^r$-functions of their arguments and
the derivatives up to order $r$ satisfy estimates like (\ref{eq.
1.13}), more precisely, for any multi-indices ${\cal A}_n=\left\{
\alpha _1,...,\alpha _n,\beta \right\} $ with $\left| {\cal
A}_n\right| =\sum\limits_{i=1}^n\left| \alpha _i\right| +\beta \leq
r$, where $ \alpha _i=\left( \alpha _i^1,...,\alpha _i^d\right) $ ,
$\beta =\left(
\beta ^1,...,\beta ^d\right) $, there exist constants $C\left( {\cal A}%
_n\right) $, $\tilde C\left( {\cal A}_n\right) $, such that, for any
$n\geq 2$ and for all $z$ in the half-plane the following inequalities
hold:
\begin{equation}
\label{eq. 2.14}
\begin{array}{c}
\left| \partial ^{
{\cal A}_{n-1}}b_n\left( z;q_1,...,q_{n-1};q\right) \right| \leq \alpha
\left( \lambda _2(p)-\kappa \right) ^{-1}\cdot C\left( {\cal A}_{n-1}\right)
\cdot \prod\limits_{i=1}^{n-1}h\left( q_i\right) \\ \left| \partial ^{{\cal A%
}_n}d_n\left( z;q_1,...,q_n;q\right) \right| \leq \alpha \left( \lambda
_2(p)-\kappa \right) ^{-1}\cdot C\left( {\cal A}_n\right) \cdot h\left(
q\right) \prod\limits_{i=1}^nh\left( q_i\right)
\end{array}
\end{equation}
\begin{equation}
\label{eq. 2.15}
\begin{array}{c}
\left| \frac d{dz}\partial ^{
{\cal A}_{n-1}}b_n\left( z;q_1,...,q_{n-1};q\right) \right| \leq \alpha
\left( \lambda _2(p)-\kappa \right) ^{-2}\cdot \tilde C\left( {\cal A}%
_{n-1}\right) \cdot \prod\limits_{i=1}^{n-1}h\left( q_i\right) \\ \left|
\frac d{dz}\partial ^{{\cal A}_n}d_n\left( z;q_1,...,q_n;q\right) \right|
\leq \alpha \left( \lambda _2(p)-\kappa \right) ^{-2}\cdot \tilde C\left(
{\cal A}_n\right) \cdot h\left( q\right) \prod\limits_{i=1}^nh\left(
q_i\right)
\end{array}
\end{equation}
where $\partial ^{{\cal A}_n}=\left( \prod\limits_{i=1}^n\partial
_{q_i}^{\alpha _i}\right) \partial _q^\beta $.
The vector $F_2^0\left( z;f_1\right) $ given by equation (\ref
{eq. 2.8}), having as coefficient functions the components $b_n,d_n$
of $\mu \left( z\right) $, belongs to ${\cal H}^{(\geq 2)}$ and is
the unique solution of equation (\ref{eq. 2.1}) for $L_2=0$.
\end{lemma}
\textit{Proof}: Suppose $\mu \in {\cal M}$, $\left\| \mu \right\| =1$, i.e. $%
\left( b_n\right) _{n\geq 2},\left( d_n\right) _{n\geq 2}$ satisfy
the estimates (\ref{eq. 2.10}) with $M=1$. Then, $\Gamma (z)\mu
=\hat \mu $ of components (\ref{eq. 2.9b}),(\ref{eq. 2.9d})
satisfies the same estimates with
$$
\hat A=\alpha \cdot \max \limits_{n\geq 2}\frac{n+1+\left\| h\right\| ^2}{%
(n-2)c_o+\lambda _2(p)-\kappa }=\frac{\alpha \left( 3+\left\| h\right\|
^2\right) }{\lambda _2(p)-\kappa }.
$$
Therefore, $\left\| \Gamma (z)\right\| <1/2$ for $\alpha $ sufficiently
small. The estimate of $\left\| \mu _0\right\| $ is immediate, therefore $%
\left\| \mu (z)\right\| \leq 2\alpha \left( \lambda _2(p)-\kappa \right)
^{-1}$.

So, we are left with the proof of the smoothness of the
coefficient functions, equations (\ref{eq. 2.14}), (\ref{eq. 2.15}).
We shall consider only the first
derivatives, i.e. $\left| {\cal A}_n\right| =1$. Consider the subspace $%
{\cal M}_1\subset {\cal M}$ of all $\mu $ with differentiable components $%
\left( b_n\right) _{n\geq 2},\left( d_n\right) _{n\geq 2}$ for which there
exists $M_1>0$, such that%
$$
\max \left\{ \max \limits_{1\leq i\leq n-1}\left| \nabla
_{q_i}b_n\left( q_1,...,q_{n-1};q\right) \right| ,\left| \nabla
_qb_n\left( q_1,...,q_{n-1};q\right) \right| \right\} \leq M_1
\prod\limits_{i=1}^{n-1}h\left( q_i\right) ,
$$
\begin{equation}
\label{eq. 2.16}\max \left\{ \max \limits_{1\leq i\leq n}\left|
\nabla _{q_i}d_n\left( q_1,...,q_n;q\right) \right| ,\left| \nabla
_qd_n\left( q_1,...,q_n;q\right) \right| \right\} \leq M_1
h(q)\prod\limits_{i=1}^nh\left( q_i\right) ,
\end{equation}
which is a Banach space with norm $\left\| \mu \right\| _1=\max
\left\{ \left\| \mu \right\| ,\inf M_1\right\} $, where $\inf $ is
taken over all $ M_1$ fulfilling (\ref{eq. 2.16}). We show that, for
small $\alpha $, $\Gamma (z)$ is a contraction in ${\cal M}_1$, as
well. Taking derivatives with respect, say, to $q_1$ in equation
(\ref{eq. 2.9b}), one obtains
$$
\begin{array}{c}
\nabla _{q_1}\hat b_n\left( q_1,...,q_{n-1};q_n\right)
\hfill \\ =-\alpha \left( e_{n,p}^0\left( q_1,...,q_n\right) -z\right)
^{-2}\nabla _{q_1}e_{n,p}^0\left( q_1,...,q_{n-1},q_n\right)
\hfill \\
\begin{array}{l}
\times \left[ \sum\limits_{i=1}^{n-1}c\left(
p-\sum\limits_{j=1}^nq_j;q_i\right) b_{n-1}\left( ...,\check
q_i,...;q_n\right) \right. \\
\left. +\int \bar c\left( p-\sum\limits_{j=1}^nq_j-q;q\right) b_{n+1}\left(
q_1,...,q_{n-1},q;q_n\right) dq\right]
\end{array}
\end{array}
\;\;\;\;\;\;\;\;\;\;\;\;\;\;\;\;\;\;\;\;\;
$$
\begin{equation}
\label{eq. 2.17}
\begin{array}[b]{r}
+\alpha \left( e_{n,p}^0\left( q_1,...,q_n\right) -z\right) ^{-1}
\hfill \\
\begin{array}{l}
\cdot \left[ \left( \nabla _qc\left( p-\sum\limits_{j=1}^nq_j;q_1\right)
-\sum\limits_{i=1}^{n-1}\nabla _pc\left( p-\sum\limits_{j=1}^nq_j;q_i\right)
\right) b_{n-1}\left( ...\check q_i...;q_n\right) \right. \\
-\int \nabla _p\bar c\left( p-\sum\limits_{j=1}^nq_j-q;q\right)
b_{n+1}\left( q_1,...,q_{n-1},q;q_n\right) dq
\end{array}
\\
\begin{array}{l}
+\sum\limits_{i=2}^{n-1}c\left( p-\sum\limits_{j=1}^nq_j;q_i\right) \nabla
_{q_1}b_{n-1}\left( q_1,...,\check q_i..,q_{n-1};q_n\right) \\
\left. +\int \bar c\left( p-\sum\limits_{j=1}^nq_j-q;q\right) \nabla
_{q_1}b_{n+1}\left( q_1,...,q_{n-1},q;q_n\right) dq\right] .
\end{array}
\hfill
\end{array}
\end{equation}
Here, $\nabla _pc$ and $\nabla _qc$ denote the gradient of the function $%
c(p,q)$ with respect to the first, respectively the second,
argument. Similar expressions are obtained for $\nabla _{q_i}\hat
d_n$, $\nabla _{q_n}\hat b_n$, and $\nabla _q\hat d_n$. Suppose that
$\left\| \mu \right\| _1=1$. Then, using the simple estimate
\begin{equation}
\label{eq. 2.18}\left| \frac{\nabla _{q_1}e_{n,p}^0\left(
q_1,...,q_{n-1},q_n\right) }{e_{n,p}^0\left( q_1,...,q_n\right) -z}\right|
\leq \bar R,
\end{equation}
where $\bar R$ is a constant independent of $n$, and also the assumption (%
\ref{eq. 1.13}), one obtains
\begin{equation}
\label{eq. 2.19}\left\| \hat \mu \right\| _1\leq \frac{\alpha \left(
a+\left\| h\right\| ^2\right) \left( \bar R+\bar C+b\right) }{\lambda
_2(p)-\kappa },
\end{equation}
where $a$ and $b$ are absolute constants and $\bar C=\max \left\{
\left| \nabla _pc(p,q)\right| ,\left| \nabla _qc(p,q)\right|
\right\} $. Equation (\ref {eq. 2.19}) shows that $\Gamma \left(
z\right) $ leaves ${\cal M}_1$ invariant and that $\left\|
\Gamma (z)\right\| <1/2$ for $\alpha $ sufficiently small.
Since $\mu _0\in {\cal M}_1$, and
$$\left\| \mu _0\right\|
_1<2\alpha \left( \lambda _2(p)-\kappa \right) ^{-1}\max \left\{
\left( \bar R+\bar C\right) ,1\right\} ,$$
we see that the solution
$\mu \left( z\right) $ of
equation (\ref{eq. 2.12}) belongs to ${\cal M}_1$ and has norm of the order of $%
\alpha /\left( \lambda _2(p)-\kappa \right) $. This finishes the
proof of the inequalities (\ref{eq. 2.14}) in the case $r=1$. The
higher values of $r$ can be treated similarly,
with stronger limitations on $\alpha $.

Finally, $\mu \left( z\right) $ and its derivatives are analytic in
the half-plane $\xi \leq \kappa ^{\prime }$ for any $\kappa ^{\prime
}\in \left( \kappa ,\lambda _2(p)\right) $ and satisfy there
inequalities like (\ref {eq. 2.14}), implying (\ref{eq. 2.15}) in
$\xi \leq \kappa $. The lemma is proved. \hfill $\square$

Finally, going back to the system (\ref{eq. 1.20})\ with $L_2=0$, we
remark that the solution $F_2^0\left( z;f_1\right) $ of the second
equation enters the first equation only through its first ($n=2$)
component, $f_2^0\left( z;f_1\right) $, which, in view of equation
(\ref{eq. 2.8}) has the form:
\begin{equation}
\label{eq. 2.20}
\begin{array}{r}
f_2^0\left( z;f_1;q_1,q_2\right) =b_2\left( z;q_2;q_1\right) f_1\left( q_1\right)
+b_2\left( z;q_1;q_2\right) f_1\left( q_2\right) \\ +\int d_2\left( z;q_1,q_2;q\right)
f_1\left( q\right) dq.
\end{array}
\end{equation}
Inserting this representation into the first equation (\ref{eq. 1.20}) and
using the notations:
\begin{equation}
\label{eq. 2.21}m_p\left( z;q\right) =\alpha \int \overline{c\left( p-q-q^{\prime
};q^{\prime }\right) }b_2\left( z;q^{\prime };q\right) dq^{\prime },
\end{equation}

\begin{equation}
\label{eq. 2.22}
\begin{array}{r}
D_p\left( z;q,q^{\prime }\right) =\frac 1\alpha \left[ c\left( p-q-q^{\prime
},q^{\prime }\right) b_2\left( z;q;q^{\prime }\right) \right. \\
\left. +\int \overline{c\left( p-q^{\prime }-q^{\prime \prime },q^{\prime
\prime }\right) }d_2\left( z;q,q^{\prime \prime };q^{\prime }\right)
dq^{\prime \prime }\right] ,
\end{array}
\end{equation}
one arrives at the following system of equations for the $n=0,1$ components:
\begin{equation}
\label{eq. 2.23}\left\{
\begin{array}{lll}
\left( e_{0,p}^0-z\right) f_0 & +\alpha \int \overline{c\left( p-q,q\right) }%
f_1\left( q\right) dq & =l_0 \\
\alpha c\left( p-q,q\right) f_0 & +\left[ a_p\left( z;q\right) -z\right]
f_1\left( q\right) +\alpha ^2\int D_p\left( z;q,q^{\prime }\right) f_1\left(
q^{\prime }\right) dq^{\prime } & =l_1
\end{array}
\right.
\end{equation}
where%

\begin{equation}\label{a-m-relation}
  a_p\left( z;q\right) =e_{1,p}^0\left( q\right) +m_p\left( z;q\right)
\end{equation}
\begin{corollary}\label{C3.3}  For $z$ \ real, the function $a_p\left(
z;q\right) $ \ is real and the kernel $D_p\left( z;q,q^{\prime
}\right) $ \ is self-adjoint.\end{corollary}

Indeed, the operator $V(z)$ defined by
$$\left( V(z)f_1\right) \left(
q\right) =m_p\left( z;q\right) f_1\left( q\right) +\alpha ^2\int
D_p\left( z;q,q^{\prime }\right) f_1\left( q^{\prime }\right)
dq^{\prime }$$
is equal to  $-A_{12}\left( A_{22}-zI\right)
^{-1}A_{21}$ appearing in equation (\ref{eq. 1.24}), which is
manifestly self-adjoint for real $z$.\\
\begin{corollary}\label{C3.4} The following asymptotic formulae hold:
\begin{equation}
\label{eq. 2.24}m_p\left( z;q\right) =-\alpha ^2\int \frac{\left| c\left(
p-q-q^{\prime };q^{\prime }\right) \right| ^2}{e_{2,p}^0\left( q,q^{\prime }\right)
-z}dq^{\prime }+O\left( \alpha ^3\right) ,
\end{equation}
where $O\left( \alpha ^3\right) $ \ is a $C^1$-function of norm
$\left\| O\left( \alpha ^3\right) \right\| _1\leq C\alpha ^3$ \ for
some constant $C$ \ depending on $\kappa $;
\begin{equation}
\label{eq. 2.25}D_p\left( z;q,q^{\prime }\right) =-\frac{\overline{c\left(
p-q-q^{\prime };q^{\prime }\right) }c\left( p-q-q^{\prime };q\right) }{%
e_{2,p}^0\left( q,q^{\prime }\right) -z}+ O\left( \alpha \right) ,
\end{equation}
where $O\left( \alpha \right) $ \ is a smooth function bounded by %
$C\, \alpha \, h(q)h(q^{\prime })$ \ for some constant $C$ \
depending on $\kappa $ . \end{corollary}

As a consequence of Lemma 3.2, the
solution $\mu \left( z\right) $ has a series representation
$\sum\limits_{n=0}^\infty \left( -\Gamma \left( z\right) \right)
^n\mu _0$ convergent in ${\cal M}_1$, the $n$-th term of which is of
the order $\alpha ^n$, wherefrom the assertion.

\section{Study of the reduced system (3.25)}
\setcounter{equation}{0} \renewcommand{\theequation}{\arabic{section}.%
\arabic{equation}} \setcounter{theorem}{0}
\subsection{The generalized Friedrichs model (a digression)}\label{sec:fried}

We collect here the needed information about the spectral
representation of the generalized Friedrichs operator $A$ acting in $%
{\cal H}^{(\leq 1)}={\mathbb C}\oplus L^2({\mathbb R}^d,dq) $, equation (\ref {eq.
1.26}). We shall study $A$ as a perturbation of $A_0=A\left( \alpha =0\right) $, so
$\alpha >0$ is supposed sufficiently small to ensure the convergence.

In order to calculate the resolvent $R_A(z)$ of $A$, one has to solve
\begin{equation}
\label{eq. 3.1}
\begin{array}[b]{r}
\left\{
\begin{array}{r}
\left( e^{(0)}-z\right) f_0 +\alpha \int \bar v\left( q\right)
f_1\left( q\right) dq=g_0 \\ \alpha v\left( q\right) f_0+\left(
a\left( q\right) -z\right) f_1\left( q\right) +\alpha ^2\int D\left(
q,q^{\prime }\right) f_1\left( q^{\prime }\right) dq^{\prime }=g_1
\end{array}
\right. \\
\end{array}
\end{equation}
for all $\left( g_0,g_1\right) =G\in {\cal H}^{(\leq
1)}$.\\
To this end the following assumptions are made:
\begin{enumerate}
\item  \label{cond.1} $a\left( q\right) $ is a real, sufficiently smooth
function, and there exist constants $C_1,C_2,C_3$, such that
\begin{equation}
\label{eq. 3.2}
\begin{array}{c}
C_1\leq a\left( q\right) \leq C_2\left| q\right| ^2+C_3\,, \\
\left| \nabla a\left( q\right) \right| \leq C_2\left( \left| q\right|
+1\right)\,, \\
\left| \partial ^\alpha a\left( q\right) \right| \leq C_2,\left|
\alpha \right| \geq 2\,;
\end{array}
\end{equation}
$a(q)$ has a unique nondegenerate minimum $\bar a$ at $\bar q_0$ and no other critical
points. We denote $I=\left[ \bar a,\infty \right) \subset {\mathbb R}$ the range of
the function $a$.

\item  \label{cond.2} The function $v\left( q\right) $ is continuous and $\left| v\left(
q\right) \right| \leq h(q)$, for some bounded, rapidly decreasing, positive $%
h$;

\item  \label{cond.3}$a\left( q\right) $ restricted to the $0$-level of $v$,
$\left\{ q:v(q)=0\right\} $, is not constant;

\item  \label{cond.4} The kernel $D\left( q,q^{\prime }\right) $ is sufficiently smooth
and there exists a constant $N$, such that, for any multi-indices
$\alpha ,\beta $ with $\left| \alpha \right| ,\left| \beta
\right| \leq r=\left[
d/2\right] +2$%
\begin{equation}
\label{eq. 3.3}\left| \partial _q^\alpha \partial _{q^{\prime }}^\beta
D\left( q,q^{\prime }\right) \right| \leq Nh(q)h(q^{\prime }).
\end{equation}
\end{enumerate}
In solving equation (\ref{eq. 3.1}) we proceed like outlined in
Section \ref{outline} , i.e. we solve the second equation
for $f_1$ in terms of $f_0$ and plug the solution in the first
equation.

Let $B$ be the operator defined on its maximal domain in $L^2\left( {\mathbb R}%
^d,dq\right) $ by the formula:
\begin{equation}
\label{eq. 3.4}Bf(q)=a\left( q\right) f\left( q\right) +\alpha ^2\int
D\left( q,q^{\prime }\right) f\left( q^{\prime }\right) dq^{\prime }.
\end{equation}
We need its resolvent $R_B(z)=\left( B-zI\right) ^{-1}$. We denote by ${\cal B}%
_r$ the Banach space of all kernels $D\left( q,q^{\prime }\right) $
satisfying condition \ref{cond.4}, i.e. which are $r$ times differentable
and satisfy (\ref{eq. 3.3}) for some $N$, endowed with the norm $\left\|
D\right\| _r=\inf N$, where the infimum is taken over all $N$ for which (\ref
{eq. 3.3}) holds.

\begin{lemma}\label{L4.1} For $\alpha $ sufficiently small
and $z\notin I $ the resolvent $R_B(z)=\left( B-zI\right) ^{-1}$ has
the form
\begin{equation}
\label{eq. 3.5}
\begin{array}[b]{l}
\left( R_B(z)g\right) \left( q\right) \\ =\left( a\left( q\right) -z\right)
^{-1}\left[ g\left( q\right) +\alpha ^2\int K\left( \alpha ,z;q,q^{\prime }\right)
g\left( q^{\prime }\right) \left( a\left( q^{\prime }\right) -z\right) ^{-1}dq^{\prime
}\right] , \\ \hfill g\in L^2\left( {\mathbb R}^d,dq\right) ,
\end{array}
\end{equation}
where the kernel $K\left( \alpha ,z;\cdot ,\cdot \right) \in {\cal B}%
_r $ and its norm is bounded for $z\in {\mathbb C}\setminus I$.
Moreover, $K$ is a ${\cal B}_r$-valued analytic function of $z$  on
${\mathbb C}\setminus I$ and its boundary values
\begin{equation}
\label{eq. 3.6}K^{\pm }\left( \alpha ,x;q,q^{\prime }\right) =\lim
\limits_{\varepsilon \searrow 0}K\left( \alpha ,x\pm i\varepsilon
;q,q^{\prime }\right)
\end{equation}
exist in ${\cal B}_r$ \ for all $x\in I$. Also, $K^{\pm }\left(
\alpha ,x;\cdot ,\cdot \right) $ are $\left[(d-1)/ 2\right]
-1$ times differentiable as a ${\cal B}_r$-valued function of $x\in
I$ and their last derivative with respect to $x$ is H\"older
continuous of exponent $\gamma =1/3$ (actually of any $\gamma <1/2$ %
for even $d$ and any $\gamma <1$ for odd $d$).
\end{lemma}
\begin{remark}\label{R4.1} We shall express the last property of $K^{\pm }$ by saying
that $I\ni x\longmapsto K^{\pm }\left( \alpha ,x;\cdot ,\cdot
\right) \in {\cal B}_r$ is $s+1/3$ times differentiable, where we
put $s=\left[
(d-1)/2\right] -1$. In order to prove it, one has to find a constant $%
\tilde N$, such that for all multi-indices $\alpha ,\beta $ with
$\left|
\alpha \right| ,\left| \beta \right| \leq r=\left[d/2\right] +2$ and $%
k=0,1,...,s$:
\begin{equation}
\label{eq. 3.7}\left| \partial _x^k\partial _q^\alpha \partial _{q^{\prime
}}^\beta K^{\pm }\left( \alpha ,x;q,q^{\prime }\right) \right| \leq \tilde
Nh(q)h(q^{\prime }),
\end{equation}
and
\begin{equation}
\label{eq. 3.8}\max \limits_{x,y\in I;\left| x-y\right| \leq 1}\frac{\left|
\partial _x^s\partial _q^\alpha \partial _{q^{\prime }}^\beta K^{\pm }\left(
\alpha ,x;q,q^{\prime }\right) -\partial _x^s\partial _q^\alpha \partial
_{q^{\prime }}^\beta K^{\pm }\left( \alpha ,y;q,q^{\prime }\right) \right| }{%
\left| x-y\right| ^{1/3}}\leq \tilde Nh(q)h(q^{\prime }).
\end{equation}
\end{remark}
\textit{Proof of Lemma \ref{L4.1}}: Let $B_0=B(\alpha =0)$, i.e. the operator
of multiplication with $a(q)$ and $D$ the integral operator of kernel $D\left(
q,q^{\prime }\right) $. Then, denoting
\begin{equation}
\label{eq. 3.9}M=\alpha ^2D\left( B_0-z\right) ^{-1}
\end{equation}
which is an integral operator of kernel $\alpha ^2D\left( q,q^{\prime }\right) \left(
a(q^{\prime })-z\right) ^{-1}$, we have, formally, the expansion:
\begin{equation}
\label{eq. 3.10}
\begin{array}{r}
R_B(z)=\left( B_0-z\right) ^{-1}\left( I+M\right) ^{-1}=\left( B_0-z\right)
^{-1} \\
+\sum\limits_{n=1}^\infty (-1)^n\left( B_0-z\right) ^{-1}M^n,
\end{array}
\end{equation}
where $\left( B_0-z\right) ^{-1}M^n$, $n\geq 1$, are integral operators of
kernels
$$
\left( a\left( q\right) -z\right) ^{-1}L_n\left( \alpha ,z;q,q^{\prime
}\right) \left( a\left( q^{\prime }\right) -z\right) ^{-1},
$$
with
\begin{equation}
\label{eq. 3.11}L_n\left( \alpha ,z;q,q^{\prime }\right) =\alpha ^{2n}\int
...\int \frac{D\left( q,q_1\right) D\left( q_1,q_2\right) ...D\left(
q_{n-1},q^{\prime }\right) }{\prod\limits_{i=1}^{n-1}\left( a\left(
q_i\right) -z\right) }dq_1...dq_{n-1}.
\end{equation}

We shall prove the convergence of the series (\ref{eq. 3.10}) in ${\cal B}_r$ and,
hence, show that $K$ satisfies all the assertions of the Lemma, by checking (by
induction) the following properties of the function (\ref{eq. 3.11}): \\ (i) $L_n\left(
\alpha ,z;\cdot ,\cdot \right) \in {\cal B}_r$ and
\begin{equation}
\label{eq. 3.12}\left\| L_n\left( \alpha ,z;\cdot ,\cdot \right) \right\|
_r\leq \left( C\alpha ^2\right) ^{n-1},
\end{equation}
where $C$ is a constant (to be specified later);\\
(ii) The limits
\begin{equation}
\label{eq. 3.13}\lim \limits_{\varepsilon \searrow 0}L_n\left( \alpha ,x\pm
i\varepsilon ;q,q^{\prime }\right) =L_n^{\pm }\left( \alpha ,x;q,q^{\prime
}\right)
\end{equation}
exist in ${\cal B}_r$ for all $x\in I$;\\
(iii) $L_n^{\pm }\left( \alpha ,x;\cdot ,\cdot \right) $ are $s+
1/3$
times differentiable, thereby they satisfy the estimates (\ref{eq. 3.7}), (%
\ref{eq. 3.8}) with $\tilde N=\left( C\alpha ^2\right) ^{n-1}$.

Indeed, for $n=1$, i.e. for $D\left( q,q^{\prime }\right) $ these
assertions hold obviously. For  $\mathfrak{Im} \,z\geq 0$, we
represent
\begin{equation}
\label{eq. 3.14}
\begin{array}[b]{r}
L_{n+1}\left( \alpha ,z;q,q^{\prime }\right) =\alpha ^2\int D\left( q,\bar
q\right) L_n\left( \alpha ,z;\bar q,q^{\prime }\right) \left( a\left( \bar
q\right) -z\right) ^{-1}d\bar q \\
\\
=i\alpha ^2\int_0^\infty dte^{izt}\int D\left( q,\bar q\right) L_n\left(
\alpha ,z;\bar q,q^{\prime }\right) e^{-ita\left( \bar q\right) }d\bar q,
\end{array}
\end{equation}
wherefrom
\begin{equation}
\label{eq. 3.15}
\begin{array}{l}
\partial _z^k\partial _q^\alpha \partial _{q^{\prime }}^\beta L_{n+1}\left(
\alpha ,z;q,q^{\prime }\right) = \\
\\
i\alpha ^2\int_0^\infty dt\left( it\right) ^ke^{izt}\int \partial _q^\alpha
D\left( q,\bar q\right) \partial _{q^{\prime }}^\beta L_n\left( \alpha
,z;\bar q,q^{\prime }\right) e^{-ita\left( \bar q\right) }d\bar q.
\end{array}
\end{equation}
Using (\ref{eq. 3.3}), the induction hypothesis and the
\textit{condition} \ref {cond.1} for $a(q)$, the internal
integral can be represented by the stationary phase method as
\begin{equation}
\label{eq. 3.16}\hat C\frac{\partial _q^\alpha D\left( q,\bar q_0\right)
\partial _{q^{\prime }}^\beta L_n\left( \alpha ,z;\bar q_0,q^{\prime
}\right) e^{-ita\left( \bar q_0\right) }}{t^{d/2}+1}+\Delta _{\alpha \beta
}\left( t;q,q^{\prime }\right) ,
\end{equation}
where $\hat C$ is an absolute constant, and the kernel $\Delta _{\alpha \beta }$ is bounded by
\begin{equation}
\label{eq. 3.17}\left| \Delta _{\alpha \beta }\left( t;q,q^{\prime }\right)
\right| \leq \bar N\left( C\alpha ^2\right) ^{n-1}\left\| h\right\| _{L_2}^2%
\frac{h(q)h(q^{\prime })}{t^{d/2+1}+1}
\end{equation}
with some constant $\bar N$ dependent on $d$ and on the function $a$. The
integral%
$$ \int_0^\infty dt\frac{\left( it\right) ^k}{t^{d/2}+1}e^{i(z-a\left( \bar q_0\right)
)t} $$ is absolutely convergent for all $k\leq s$ and defines a
continuous function of $z$ in $\mathfrak{Im}\,z\geq 0$, which, for
$k=s$, is H\"older continuous with respect to this variable. We have
that the contribution to (\ref{eq. 3.15}) of the first term in
(\ref{eq. 3.16}) has the estimate
\begin{equation}
\label{eq. 3.18}
\begin{array}{r}
\left| i\alpha ^2\hat C\int_0^\infty dt \left( it\right)
^ke^{it(z-a\left( \bar q_0\right) )}(t^{d/2}+1)^{-1}
\partial _q^\alpha D\left( q,\bar q_0\right) \partial _{q^{\prime }}^\beta
L_n\left( \alpha ,z;\bar q_0,q^{\prime }\right) \right| \\ \leq {\tilde C}h\left( \bar
q_0\right) ^2\tilde N(C\alpha ^2)^{n-1}h(q)h(q^{\prime }),
\end{array}
\end{equation}
where $\tilde C$ is a constant. One proves in the same way the
H\"older condition (\ref{eq. 3.8}) for $k=s$. A similar estimate
holds for the integral of the second  term in (\ref{eq. 3.16}):
$$
\left| \int_0^\infty \left( it\right) ^ke^{izt}\Delta _{\alpha \beta }\left(
t;q,q^{\prime }\right) dt\right| \leq \tilde C\left\| h\right\| _{L_2}^2\bar
N(C\alpha ^2)^{n-1}h(q)h(q^{\prime }).
$$
By taking $C=\max \left\{ N,\tilde C\left( \left| h\left( \bar
q_0\right) \right| ^2\tilde N+\left\| h\right\| _{L_2}^2\bar
N\right) \right\} $, one gets the estimate (\ref{eq. 3.12}), the
existence of the limit (\ref {eq. 3.13}) and the assertion (iii) for
$n$ replaced by $n+1$. \hfill $\square$\\

Once we have $R_B(z)$, it is an easy matter to write down the
solution of equation (\ref{eq. 3.1}) for $z\in {\mathbb C}\setminus
I$ as
\begin{equation}
\label{eq. 3.19}f_1=R_B(z)\left[ g_1-\alpha f_0v\right] ,
\end{equation}
where
\begin{equation}
\label{eq. 3.20}f_0=\frac 1{\Delta (z)}\left[ g_0-\alpha \left(
v,R_B(z)g_1\right) \right]
\end{equation}
whenever $\Delta (z)\neq 0$. Here,
\begin{equation}
\label{eq. 3.21}\Delta (z)=e^{\left( 0\right) }-z-\alpha ^2\left(
v,R_B(z)v\right)
\end{equation}
Clearly, $\Delta (z)$ is analytic in ${\mathbb C}\setminus I$, has
limits at the cut $I$:
\begin{equation}
\label{eq. 3.22}\lim _{\varepsilon \searrow 0}\Delta \left( x\pm
i\varepsilon \right) =\Delta ^{\pm }\left( x\right) ,x\in I
\end{equation}
and the limits $\Delta ^{\pm }\left( x\right) $ are $s+1/3$ times
differentiable, by the same reasoning as in Lemma \ref{L4.1}. More precisely,%
$$
\left| \frac{d^k}{dx^k}\left( \Delta ^{\pm }\left( x\right) +x\right)
\right| \leq const,\;k=0,...,s+1/3.
$$

As one can read from equations (\ref{eq. 3.19}), (\ref{eq.
3.20}), the continuous spectrum of the operator $A$ equals the
interval $I$. Besides, the real zeroes of $\Delta (z)$ below $\bar
a$, if any, are eigenvalues of $A$. Since $-\alpha ^2\left(
v,R_B(x)v\right) $ is decreasing for $x<\bar a$, $\Delta (x)$ is
strictly decreasing from +$\infty $ to $\Delta \left( \bar a\right)
$ on this interval, therefore $A$ has one simple eigenvalue $e<\bar
a$ with eigenvector $\psi _0=\left( f_0,f_1=-\alpha
f_0R_B(e)v\right)\in {\cal H}^{\left( \leq 1\right)}$, if, and only
if, $\Delta \left( \bar a\right) <0$. As, for small $\alpha $

$$ \pm \mathfrak{Im}\,\Delta ^{\pm }\left( x\right) =\alpha ^2\pi \int\limits_{a(q)=x}\left|
v(q)\right| ^2dq+0\left( \alpha ^4\right) >0,x\in I, $$ in view of \textit{condition}
\ref{cond.3}, it follows that there are no eigenvalues of $A$ embedded in the
continuous spectrum (see \cite{citari Fried}).
\begin{remark}\label{R4.2} It is easy to show using the explicit formulae
for $R_A(z)$ that the general criteria of the absence of the singular continuous
spectrum \cite{Reed} are met in our case, hence that the continuous spectrum $I$ is
absolutely continuous. Therefore, we have
\begin{equation}
\label{eq. 3.23}{\cal H}^{\left( \leq 1\right) }=\left\{
\begin{array}{ll}
{\cal H}^{ac}, & \Delta \left( \bar a\right) \geq 0 \\
\left\{ c\psi _0\right\} \oplus {\cal H}^{ac}, & \Delta \left( \bar a\right)
<0
\end{array}
\right.
\end{equation}
\end{remark}

We come now to the scattering theory for the pair of self-adjoint
operators $%
(A,B_0)$, where we denoted $B_0$ the operator of multiplication with
$a(q)$ acting in ${\cal H}^{\left( 1\right) }=L^2\left( {\bf
R}^d,dq\right) $. We denote $E:{\cal H}^{\left( 1\right)
}\rightarrow {\cal H}^{\left( \leq 1\right) }$ the injection
$E\varphi =\left( 0,\varphi \right) \in \mathcal{H}^{\left( \leq
1\right) }$, $\varphi \in {\cal H}^{\left( 1\right) }$. Known
existence criteria for the wave operators (see
e.g.\cite{Reed}) can be applied to our case and ensure
the existence of the strong limit:
$$
s-\lim _{t\rightarrow \infty }e^{itA}Ee^{-itB_0}=\Omega ^{+} ,
$$
which is a unitary operator $\Omega ^{+}:{\cal H}%
^{\left( 1\right) }\rightarrow {\cal H}^{ac}\subset {\cal H}^{\left(
\leq 1\right) }$.
The generalized eigenfunctions of
the operator $B_0$ are $\delta _q(\cdot )=\delta \left( q-\cdot
\right) $, therefore, using known formulae in scattering theory, one
can take
\begin{equation}
\label{eq. 3.23a}\psi ^q=\Omega ^{+}\delta _q=\lim _{\varepsilon \searrow
0}i\varepsilon R_A\left( a\left( q\right) -i\varepsilon \right) E\delta _q
\end{equation}
as generalized eigenvectors of $A$ corresponding to the
eigenvalue $a(q)$. Explicitly, in view of  (\ref{eq.
3.19}), (\ref{eq. 3.20}) and Lemma \ref{L4.1}, one gets
for $\psi ^q=\left( f_0^q,f_1^q\left( \cdot \right) \right) $
the following expressions:
\begin{eqnarray}
\label{eq. 3.24}&&f_0^q=-\frac{\alpha }{\Delta ^{-}\left( a\left(
q\right) \right) }\left[ v(q)+\alpha ^2\int \frac{K^{-}\left( \alpha
,a\left(
q\right) ;q^{\prime },q\right) \bar v\left( q^{\prime }\right) dq^{\prime }}{%
a\left( q^{\prime }\right) -a\left( q\right) +i0}\right],\\
\label{eq. 3.25} &&f_1^q\left( q^{\prime }\right) =\delta \left( q-q^{\prime }\right) +\frac{%
\alpha ^2K^{-}\left( \alpha ,a\left( q\right) ;q^{\prime },q\right) }{%
a\left( q^{\prime }\right) -a\left( q\right) +i0}-\\
\nonumber &&\alpha f_0^q\frac 1{a\left( q^{\prime }\right) -a\left(
q\right) +i0}\left[ v(q^{\prime })+\alpha ^2\int \frac{K^{-}\left(
\alpha ,a\left( q\right) ;q^{\prime },q^{\prime \prime }\right) \bar
v\left( q^{\prime \prime }\right) dq^{\prime \prime }}{a\left(
q^{\prime \prime }\right) -a\left( q\right) +i0}\right].
\end{eqnarray}

This somewhat formal derivation of the formulas
(\ref{eq. 3.24}), (\ref{eq. 3.25})
will be justified by the next lemma, which proves that $\psi ^q\in {\mathbb C}%
\oplus {\cal S}^{\prime }\left( {\mathbb R}^d\right) $ (where ${\cal S}^{\prime
}\left( {\mathbb R}^d\right) $ is the space of tempered distributions) and that it
verifies the intertwining property of the wave-operator $\Omega ^{+}$.
\begin{lemma}\label{L4.2}
Let $d\geq 3.$ Then \\ 1. For every fixed $q\in {\mathbb R}^d$, $f_0^q$ is finite and
it is bounded and continuous as a function of $q$.\\ 2. For every fixed $q\in {\bf
R}^d$, $f_1^q\left( \cdot \right) \in {\cal S}^{\prime }\left( {\mathbb R}^d\right) $;
moreover, for every fixed
$q^{\prime }\in {\mathbb R}^d$, $f_1^q\left( q^{\prime }\right) \in {\cal S%
}^{\prime }\left( {\mathbb R}^d\right) $ with respect to $q$.\\ 3. For $\varphi \in
{\cal S}\left( {\mathbb R}^d\right) $, let us consider the vector $\psi _\varphi
=\left( C_{\varphi ,0},C_{\varphi ,1}\left( \cdot \right) \right) \in {\cal H}^{\left(
\leq 1\right) }$, where
\begin{equation}
\label{eq. 3.26}C_{\varphi ,0}=\int f_0^q\varphi \left( q\right) dq,
\end{equation}
\begin{equation}
\label{eq. 3.27}C_{\varphi ,1}\left( q^{\prime }\right) =\int f_1^q\left(
q^{\prime }\right) \varphi \left( q\right) dq.
\end{equation}
Then, for any $\varphi _1,\varphi _2\in {\cal S}\left( {\mathbb R}%
^d\right) $,
\begin{equation}
\label{eq. 3.28}\left( \psi _{\varphi _1},\psi _{\varphi _2}\right) _{{\cal H%
}^{\left( \leq 1\right) }}=\bar C_{\varphi _1,0}C_{\varphi _2,0}+\int \bar
C_{\varphi _1,1}\left( q\right) C_{\varphi _2,1}\left( q\right) dq=\left(
\varphi _1,\varphi _2\right) _{L_2},
\end{equation}
therefore the application $\varphi \mapsto \psi _\varphi $ extends to an isometry
$\Omega ^{+}:L^2\left( {\mathbb R}^d,dq\right) \rightarrow {\cal H}^{\left( \leq
1\right) }$.\\ 4. The range of the operator $\Omega ^{+}$ is ${\cal H}^{ac}$ and
$A\Omega ^{+}=\Omega ^{+}B_0$.
\end{lemma}
\begin{remark}\label{R4.3}
The relation (\ref{eq. 3.28}) may be written in the following formal
way
\begin{equation}
\label{eq. 3.28a}\left( \psi ^q,\psi ^{q^{\prime }}\right) _{{\cal H}%
^{\left( \leq 1\right) }}=\bar f_0^qf_0^{q^{\prime }}+\left(
f_1^q,f_1^{q^{\prime }}\right) _{{\cal H}^{\left( 1\right) }}=\delta \left(
q-q^{\prime }\right) ,
\end{equation}
meaning the orthonormality of the generalized functions $\left\{ \psi ^q,\;q\in
{\mathbb R}^d\right\} $.
\end{remark}
\begin{remark}\label{R4.4}
Usually, the generalized eigenvectors of a self-adjoint operator $A$
acting in the Hilbert space ${\cal H}$ are defined as derivatives $
{dE_\lambda \varphi }/{d\sigma _\varphi \left( \lambda \right)}$,
where $ \varphi \in {\cal H}$ is an arbitrary vector, $\left\{
E_\lambda \right\} $ is the family of spectral projections of $A$,
and $\sigma _\varphi \left( \lambda \right) $ is the spectral
measure corresponding to $\varphi $.
Moreover, if $A$ leaves invariant a certain dense linear subspace ${\cal H}%
_{+}\subset {\cal H}$ and ${\cal H}_{+}$ has a Hilbert space
structure such that the inclusion is quasi-nuclear, then
the derivative ${dE_\lambda \varphi }/{d\sigma _\varphi
\left( \lambda \right) }=\chi _\lambda $ exists as an element of the
conjugate space ${\cal H}_{-}={\cal H}_{+}^{*}$ and it is
an eigenvector with eigenvalue $\lambda $ of the adjoint:
$\left( A\mid _{{\cal H}_{+}}\right) ^{*}$, of the restriction of
$A$ to ${\cal H}_{+}$, which is an extension of $A$. The vectors
$\chi _\lambda \in {\cal H}_{-}$ are usually called generalized
eigenvectors of the operator $A$. It can be shown that the
generalized vectors introduced above are generalized eigenvectors of
$A$ in this sense. The same remark is valid for the generalized
eigenvectors of the operator $H_p $ (which will be constructed
farther on).
\end{remark}
\textit{Proof of Lemma \ref{L4.2}}: \\
1. This assertion follows easily from the representation
\begin{eqnarray*}
&& \int K^{-}\left( \alpha ,a\left( q\right) ;q^{\prime },q\right)
\bar v\left( q^{\prime }\right) [a\left( q^{\prime }\right) -a\left(
q\right) +i0]^{-1} dq^{\prime}= \\
&& i\int\limits_0^\infty dt\int e^{it\left( a\left( q^{\prime
}\right) -a\left( q\right) \right) }K^{-}\left( \alpha ,a\left(
q\right) ;q^{\prime },q\right) \bar v\left( q^{\prime }\right)
dq^{\prime }
\end{eqnarray*}
by applying the stationary phase method as done already in the proof
of Lemma \ref{L4.1}.\\
2. In order to prove the second assertion, we have to consider
$C_{\varphi ,1}\left( q^{\prime }\right) $. To this aim, we
represent the $q^{\prime \prime }$-integral in (\ref{eq. 3.25}) as
before, using the stationary phase method:
\begin{equation}
\label{eq. 3.29}
\begin{array}{l}
I\left( x;q^{\prime }\right) :=i\int\limits_0^\infty dt\int e^{it\left(
a\left( q^{\prime \prime }\right) -x\right) }K^{-}\left( \alpha ,x;q^{\prime
},q^{\prime \prime }\right) \bar v\left( q^{\prime \prime }\right)
dq^{\prime \prime } \\
=i\int\limits_0^\infty dt\left[ \frac{\hat C}{t^{d/2}+1}e^{it\left( a\left(
\bar q_0\right) -x\right) }K^{-}\left( \alpha ,x;q^{\prime },\bar q_0\right)
\bar v\left( \bar q_0\right) +\Delta \left( x;q^{\prime },t\right) \right] ,
\end{array}
\end{equation}
where the correction term $\Delta $ satisfies the estimates%
$$
\left| \partial _x^k\partial _q^\alpha \Delta \left( x;q^{\prime },t\right)
\right| \leq \frac{\hat Nh\left( q^{\prime }\right) }{t^{d/2+1}+1}
$$
for all multi-indices $\alpha $, $\left| \alpha \right| \leq \left[
d/2\right] +1$, and $k=0,...,s+1/3$, where $\hat N$ is a constant.
Hence, $I\left( x;q^{\prime }\right) $
fulfills for $d\geq 3$ the estimates%
$$
\left| \partial _x^k\partial _q^\alpha I\left( x;q^{\prime }\right) \right|
\leq \tilde Nh\left( q^{\prime }\right) ;\,\left| \alpha \right| \leq \left[
d/2\right] +1,k=0,...,s+1/3.
$$
The contribution of this term to $C_{\varphi ,1}\left( q^{\prime }\right) $
is:%
$$
\int dq\int dq^{\prime \prime }\frac{f_0^q\varphi \left( q\right)
K^{-}\left( \alpha ,a(q);q^{\prime },q^{\prime \prime }\right) \bar v\left(
q^{\prime \prime }\right) }{\left( a\left( q^{\prime }\right) -a\left(
q\right) +i0\right) \left( a\left( q^{\prime \prime }\right) -a\left(
q\right) +i0\right) }
$$
\begin{equation}
\label{eq. 3.30}=\int dq\frac{f_0^q\varphi \left( q\right) I\left( a\left( q\right)
;q^{\prime }\right) }{a\left( q^{\prime }\right) -a\left( q\right) +i0}=\int_{{\mathbb
R}}dx\frac{m(x)I(x;q^{\prime })}{a\left( q^{\prime }\right) -x+i0},
\end{equation}
where
\begin{equation}
\label{eq. 3.31}m(x)=\int\limits_{a\left( q\right) =x}f_0^q\varphi \left(
q\right) dq.
\end{equation}
As it follows from the proof of the point 1, $m(x)$ is $s+1/3$ times
differentiable. The same property is shared by $I\left( x;q^{\prime
}\right) $ as a function of $x$ for every fixed $q^{\prime }$.
Therefore, the integral over $x$ in ($\ref{eq. 3.30}$) converges.
The convergence of the other terms entering $C_{\varphi ,1}\left(
q^{\prime }\right) $ can be proved similarly.\\
3. Using the representation%
$$ \varphi \left( q\right) =\int \varphi \left( q_0\right) \delta \left( q-q_0\right)
dq_0,\;\varphi \in {\cal S}\left( {\mathbb R}^d\right) $$ and the formula (\ref{eq.
3.23a}) we find that
\begin{equation}
\label{eq. 3.32}\Omega ^{+}\varphi =\int \varphi \left( q_0\right) \psi
^{q_0}dq_0=\left( C_{\varphi ,0},C_{\varphi ,1}\left( \cdot \right) \right) \in {\cal
H}^{ac}\subseteq {\cal H}^{\left( \leq 1\right) }.
\end{equation}
In view of the unitarity of the application $\Omega ^{+}: L^2\left({\mathbb
R}^d\right) \rightarrow {\cal H}^{ac}$, one has
\begin{equation}
\label{eq. 3.33}
\begin{array}{l}
\left( \varphi _1,\varphi _2\right) _{L^2({\mathbb R}^d) }=\left( \Omega ^{+}\varphi
_1,\Omega ^{+}\varphi _2\right)
\\ =\bar C_{\varphi _1,0}C_{\varphi _2,0}+\int \bar C_{\varphi _1,1}\left(
q\right) C_{\varphi _2,1}\left( q\right) dq.
\end{array}
\end{equation}
4. Since ${\cal S}\left( {\mathbb R}^d\right) $ is dense in $L^2\left( {\mathbb R}%
^d\right) ,$ the image $\Omega ^{+}{\cal S}\left( {\bf
R}^d\right) $ is dense in ${\cal H}^{ac}$, therefore, in view of the
unitarity of $\Omega ^{+} $, $\Omega ^{+}L^2({\bf
R}^d) ={\cal H}^{ac}$. The intertwining property $A\Omega
^{+}=\Omega ^{+}B$ is obtained in the standard way. Lemma \ref{L4.2}
is proved.  \hfill $\square$

This lemma implies in particular that any vector $\psi \in {\cal
H}^{ac}$
has a unique representation as%
$$ \psi =\psi _f=\int_{{\mathbb R}^d}f\left( q_0\right) \psi ^{q_0}dq_0:=\lim
\limits_{\varphi _n\rightarrow f}\psi _{\varphi _n}\,\,,\,\,\,f\in L^2\left( {\mathbb R}%
^d\right) .
$$
Here, the limit in the right-hand side is meant in
${\cal H}^{ac}$ and $%
\left\{ \varphi _n\right\} $ is a sequence of elements of ${\cal S}\left( {\mathbb
R}^d\right) $ converging to $f$ in $L_2.$

\subsection{Construction of the one-boson subspace}

As explained in Section \ref{outline}, the construction of the
one-boson subspace of $H_p$ relies on the spectral representation of
the operators $ \left\{A_p\left( \xi \right)\right\}_{\xi
\leq \kappa } $, see (\ref{eq. 1.25}), entering the reduced
system (\ref{eq. 2.23}):
\begin{equation}
\label{eq. 4.1}
\begin{array}{lll}
\left( A_p\left( \xi \right) F\right) _0 & =e_{0,p}^0f_0 & +\alpha \int
\overline{c\left( p-q,q\right) }f_1\left( q\right) dq \\  &  &  \\
\left( A_p\left( \xi \right) F\right) _1\left( q\right) & =\alpha c\left(
p-q,q\right) f_0 & +a_p\left( \xi ;q\right) f_1\left( q\right) \\
&  & +\alpha ^2\int D_p\left( \xi ;q,q^{\prime }\right) f_1\left( q^{\prime
}\right) dq^{\prime }, \\
&  &  \\
&  & F=\left( f_0,f_1\left( \cdot \right) \right) \in {\cal H}^{\left( \leq
1\right) }.
\end{array}
\end{equation}
Since for any $\xi \leq \kappa $ the operator $A_p\left(
\xi \right)$ satisfies all the assumptions of the previous
subsection, there exists a family
\begin{equation}
\label{eq. 4.2}\left\{F_{\xi ,1}^q=\left( f_{\xi ,0}^q, f_{\xi ,1}^q\left( \cdot
\right) \right)\right\}_{q\in {\mathbb R}^d}
\end{equation}
of generalized eigenvectors of $A_p\left( \xi \right) $ with eigenvalues $
\left\{a_p\left( \xi ;q\right)\right\}_{q\in {\mathbb R}^d}$, given by (\ref{eq.
3.24}), (\ref{eq. 3.25}),
where $a\left( q\right) $ is replaced by $a_p\left( \xi ;q\right) $, and $%
\Delta ^{-}$, $K^{-}$ by the functions $\Delta _\xi ^{-}$, $K_\xi ^{-}$,
entering the expression of the resolvent of $A_p\left( \xi \right) $.
Let $F_{\xi ,2}^q$ be constructed in terms of $f_{\xi
,1}^q\left( \cdot \right) $ according to (\ref{eq. 2.8}), i.e.
$F_{\xi ,2}^q=S\left( \xi \right) f_{\xi ,1}^q$ where the
application $S\left( \xi \right) $ was introduced in equation
(\ref{eq. 2.7a}) (more precisely, $S(\xi )$ is the extension of that
operator to the space ${\cal B}_1^{(k)}$ defined below), where the
coefficient functions are the solution of the fixed point equation
(\ref {eq. 2.12}). Then, the complete sequence
\begin{equation}
\label{eq. 4.3}F_\xi ^q=\left( F_{\xi ,1}^q,F_{\xi ,2}^q\right) =\left(
f_{\xi ,0}^q,f_{\xi ,1}^q\left( q_1\right) ,f_{\xi ,2}^q\left(
q_1,q_2\right) ,...\right)
\end{equation}
satisfies the equation
\begin{equation}
\label{eq. 4.4}H_pF_\xi ^q=\xi F_\xi ^q+\left( a_p\left( \xi ;q\right) -\xi
\right) \hat F_\xi ^q,
\end{equation}
where we denoted $\hat F_\xi ^q=\left( F_{\xi ,1}^q,0\right) $.
Therefore, if $\xi \left( q\right) $ is a solution of equation
\begin{equation}
\label{eq. 4.5}a_p\left( \xi ;q\right) -\xi =0,
\end{equation}
then $F_{\xi (q)}^q$ is a generalized eigenvector of the operator
$H_p$ with eigenvalue $\xi\left( q\right) \equiv \xi_{p} \left(
q\right) $, cf (\ref{eq. 1.27'}) in  Remark \ref{R2.1}.

We shall give below sense to the generalized eigenfunctions $F_\xi
^q$ as elements of the dual $\left( {\cal B}^{(k)}\right) ^{\prime
}$ of an auxiliary Banach space ${\cal B}^{(k)}$, $k=\left[
d/2\right] +2$, densely and continuously embedded in the
Fock space ${\cal F}$.
\begin{definition}\label{D4.1}  Let us denote ${\cal B}_n^{(k)}$ the space of all
symmetric functions $g$ of $n$ variables $q_1,...,q_n\in {\mathbb R}^d$, $k$ times
continuously differentiable with respect to each $q_i$, and for which there exists a
constant $C$ such that, for all multi-indices $\alpha =\left( \alpha _1,...,\alpha
_n\right) $, $\alpha _i=\left( \alpha _i^1,...,\alpha _i^d\right) $, $\left| \alpha
_i\right| =\sum\limits_{s=1}^d\alpha _i^s\leq k $, one has
\begin{equation}
\label{eq. 4.7}\left| \partial_{q}^\alpha g\left( q_1,...,q_n\right) \right| \leq C
\prod\limits_{i=1}^n h\left( q_i\right) ,\;\forall q_1,...,q_n \in {\mathbb R}^d.
\end{equation}
It is a Banach space if endowed with the norm $\left\|
g\right\| _n^{(k)}=\inf C$, where the infimum is taken over all $C$
for which the estimate (\ref{eq. 4.7}) holds.
\end{definition}
Clearly, the inclusion ${\cal B}_n^{(k)}\subset {\cal H%
}^{(n)}$ is continuous and ${\cal B}_n^{(k)}$ is dense in ${\cal H}^{(n)}$.
Let next
\begin{equation}
\label{eq. 4.8}{\cal B}^{(k)}={\mathbb{C}}+{\cal B}_1^{(k)}+...+{\cal B}%
_n^{(k)}+...\subset {\cal F}
\end{equation}
be the space of sequences%
$$
G=\left( g_0,g_1\left( q_1\right) ,...,g_n\left( q_1,...,q_n\right)
,...\right) ,\;g_0\in {\mathbb C},g_n\in {\cal B}_n^{(k)},
$$
with norm
\begin{equation}
\label{eq. 4.9}\left\| G\right\| _{{\cal B}^{(k)}}^{(k)}=\left[ \left|
g_0\right| ^2+\sum\limits_{n\geq 1}\frac 1{n!}\left( \left\| g_n\right\|
_n^{(k)}\right) ^2\right] ^{1/2}.
\end{equation}
Obviously, ${\cal B}^{(k)}$ is continuously and densely embedded in
the Fock space ${\cal F}
$, as required. The dual $\left( {\cal B}^{(k)}\right) ^{\prime }$ of ${\cal %
B}^{(k)}$ consists of sequences $F=\left( f_0,f_1,...,f_n,...\right)
$, where $f_0\in {\mathbb C}$, and $f_n\in \left( {\cal
B}_n^{(k)}\right) ^{\prime } $ are linear continuous functionals on
${\cal B}_n^{(k)}$; thereby, the value of $F$ at an element $G\in
{\cal B}^{(k)}$ is given by the series:
\begin{equation}
\label{eq. 4.10}\left( F,G\right) =\left[ \bar f_0g_0+\sum\limits_{n\geq
1}\frac 1{n!}\left( f_n,g_n\right) \right] ^{1/2},
\end{equation}
and the norm of $F$ is
\begin{equation}
\label{eq. 4.11}\left\| F\right\| _{\left( {\cal B}^{(k)}\right) ^{\prime
}}^{(k)}=\left[ \left| f_0\right| ^2+\sum\limits_{n\geq 1}\frac 1{n!}\left(
\left\| f_n\right\| _{\left( {\cal B}_n^{(k)}\right) ^{\prime }}\right)
^2\right] ^{1/2}.
\end{equation}
Clearly, ${\cal F}\subset \left( {\cal B}^{(k)}\right) ^{\prime }$
and the inclusion is continuous.
\begin{lemma} \label{4.3}  For every $q\in {\mathbb R}^d$ \ and $\xi \leq
\kappa $ , $F_\xi ^q\in \left( {\cal B}^{(k)}\right) ^{\prime }$ \
and has the representation
\begin{equation}
\label{eq. 4.12}F_\xi ^q=\hat \delta _q+\tilde F_\xi ^q,
\end{equation}
where $\hat \delta _q=\left( 0,\delta _q,0,...\right) $ \ and
\begin{equation}
\label{eq. 4.13}\left\| \tilde F_\xi ^q\right\| _{\left( {\cal B}%
^{(k)}\right) ^{\prime }}\leq M\alpha \,\, h\left( q\right)
\end{equation}
for some constant $M$.
\end{lemma}
\textit{Proof}: We prove this statement in three steps.\\
I. The $n=0$ component of $\tilde{F}_\xi ^q$, $\tilde{f}_{\xi
,0}^q$, is given by equation (\ref{eq. 3.24}), where
$v(q)=c(p-q,q)$, $a(q)=a_p\left( \xi ,q\right) $ and $K^{-}=K_\xi
^{-}$. If \textit{condition} \ref{cond.2} in Section \ref
{sec:fried} is fulfilled for every $\xi \leq \kappa $ and $p$, we have $%
\Delta ^{-}\left( a\left( q\right) \right) \geq \tau >0$, therefore we
obtain for the first term in (\ref{eq. 3.24})
\begin{equation}
\label{eq. 4.14}\left|\frac{(-\alpha v(q))}{\Delta ^{-}\left(
a\left( q\right) \right) }\right| \leq \frac \alpha \tau h(q).
\end{equation}
The second term in (\ref{eq. 3.24}) is treated using as before the
stationary phase method, which gives
\begin{equation}
\label{eq. 4.15}
\begin{array}[b]{l}
\int K^{-}\left( \alpha ,a\left( q\right) ;q^{\prime },q\right) \bar v\left(
q^{\prime }\right) \left( a\left( q^{\prime }\right) -a\left( q\right)
+i0\right) ^{-1}dq^{\prime } \\
=i\int\limits_0^\infty dt\left[ \hat C\left( t^{d/2}+1\right)
^{-1}e^{it\left( a\left( \bar q_0\right) -a\left( q\right) \right)
}K^{-}\left( \alpha ,a\left( q\right) ;\bar q_0,q\right) \bar v\left( \bar
q_0\right) +\Delta \left( q,t\right) \right] ,
\end{array}
\end{equation}
where
\begin{equation}
\label{eq. 4.16}\left| \Delta \left( q,t\right) \right| \leq C
h(q)\left( t^{d/2+1}+1\right) ^{-1}.
\end{equation}
for some constant $C$. Equations (\ref{eq. 4.14}),
(\ref{eq. 4.15}), (\ref{eq. 4.16}) provide
\begin{equation}
\label{eq. 4.17}\left| \tilde{f}_{\xi ,0}^q\right| \leq B\alpha
\,\,h(q).
\end{equation}
II. The $n=1$ component of $F_\xi ^q$, $f_{\xi ,1}^q=\delta _q+\tilde f_{\xi ,1}^q$, is
given by equation (\ref{eq. 3.25}), with the same assignments for $v$,
$a$, and $K^{-}$. Again, reducing the estimate of every integral entering $%
\left( \tilde f_{\xi ,1}^q,g_1\right) $ for a generic $g_1\in {\cal B}_1^{(k)}$ to the
estimate of the corresponding oscillatory integral, and using thereby the estimate
(\ref{eq. 4.14}), we obtain
\begin{equation}
\label{eq. 4.18}\left\| \tilde f_{\xi ,1}^q\right\| _{\left( {\cal B}%
_1^{(k)}\right) ^{\prime }}\leq L\alpha ^2\,\, h\left( q\right) .
\end{equation}
III. The higher components of $\tilde{F}_\xi ^q$,  $\{\tilde{f}_{\xi
,n}^q \}_{n\geq 2}$ , are estimated using their representation
(\ref{eq. 2.8}) in terms of $f_{\xi ,1}^q$. We have
\begin{equation}
\label{eq. 4.19}
\begin{array}[b]{l}
\left( \tilde{f}_{\xi ,n}^q,g_n\right) = \\ =\sum\limits_{i=1}^n\int
b_n\left( q_1,...,\check q_i..,q_n;q_i\right) f_{\xi
,1}^q\left( q_i\right) g_n\left( q_1,...,q_n\right) dq_1...dq_n \\
+\int d_n\left( q_1,...,q_n;q^{\prime }\right) f_{\xi ,1}^q\left(
q^{\prime }\right) g_n\left( q_1,...,q_n\right)
dq_1...dq_ndq^{\prime }.
\end{array}
\end{equation}
Using the estimates for $b_n$, $d_n$ and their derivatives (see
(\ref{eq. 2.10}) and Lemma \ref{L3.2}), and also the bound (\ref{eq.
4.18}), we have that
$$
\begin{array}[b]{l}
\int \left| \int b_n\left( q_1,...,\check q_i..,q_n;q_i\right)
\tilde{f}_{\xi,1}^q\left( q_i\right) g_n\left( q_1,...,q_n\right) dq_i\right|
dq_1...d\check q_i...dq_n \\ \leq C_1\alpha \left( \lambda _2(p)-\kappa \right)
^{-1}\left\| g_n\right\| _{{\cal B}_n^{(k)}}\left( 1+L\alpha ^2\right) h\left(
q\right) \left( \int h(q^{\prime })dq^{\prime }\right) ^{n-1}
\end{array}
$$
for $i=1,...,n$, and also that
$$
\begin{array}[b]{l}
\left| \int d_n\left( q_1,...,q_n;q^{\prime }\right) \tilde{f}_{\xi ,1}^q\left(
q^{\prime }\right) g_n\left( q_1,...,q_n\right) dq_1...dq_ndq\right| \\ \leq C_2\alpha
\left( \lambda _2(p)-\kappa \right) ^{-1}\left\| g_n\right\| _{{\cal B}_n^{(k)}}\left(
1+L\alpha ^2h\left( q\right) \right) h\left( q\right) \left( \int h(q^{\prime
})dq^{\prime }\right) ^n.
\end{array}
$$
Hence, with suitable constants $\tilde C$, $\tilde L$,
\begin{equation}
\label{eq. 4.20}\left\| \tilde{f}_{\xi ,n}^q\right\| _{\left( {\cal B}%
_n^{(k)}\right) ^{\prime }}\leq \tilde C\frac \alpha {\lambda
_2(p)-\kappa }\left( 1+\tilde L\alpha ^2\right) \cdot h\left(
q\right) \left( \int h(q^{\prime })dq^{\prime }\right) ^{n-1}.
\end{equation}
Putting together equations (\ref{eq. 4.17}), (\ref{eq. 4.18}) and (\ref{eq. 4.20}%
), we obtain (\ref{eq. 4.13}). The lemma is proved. \hfill $\square$



Now we come back to the study of the generalized
eigenvectors $F_{\xi (q)}^q$.

Let us remark that $a_p\left( \xi ;q\right) $ is, for every fixed $q$, a
smooth, monotonously decreasing function of $\xi $ on $\left( -\infty
,\kappa \right] $. If
\begin{equation}
\label{eq. 4.6}G_p^{(1),\kappa }=\left\{ q\in {\mathbb R}^d: \;a_p\left( \kappa
;q\right) -\kappa <0\right\} ,
\end{equation}
then equation (\ref{eq. 4.5}) has a unique solution $\xi \left( q\right) <\kappa $ if
$q\in G_p^{(1),\kappa }$, and no solution if $q\notin G_p^{(1),\kappa }$, see Figure
1.

\begin{proposition}\label{Rnew}
The function $\xi \left( q\right) \equiv \xi _p\left( q\right) $ can be represented in
the form
\begin{equation}
\label{eq. 4.55a}\xi _p\left( q\right) =\varepsilon \left( q\right) +\gamma \left(
p-q\right)
\end{equation}
where the function $\gamma \left( k\right) $ is defined in the domain $\{
k: p-k\in G_p^{(1),\kappa }\} $.
\end{proposition}
\textit{Proof}: Indeed, let us use the expansion
\begin{equation}\label{4.55b}
\mu =\mu _0-\Gamma \mu _0+\Gamma ^2\mu _0+...
\end{equation}
for the solution of equation (\ref{eq. 2.12}), and remark that the function
$b_2^0\left( q_1;q\right) $ appearing in (\ref{eq. 2.13}) can be written in the form
\begin{equation}\label{4.55c}
  b_2^0\left( q_1;q\right) \equiv b_{2,p}^0\left( q_1;q\right) =\varphi ^0_{q_1}
  \left( p-q;z-\varepsilon \left( q\right) \right)
\end{equation}
with $\varphi ^0_{q_1}\left( k;w\right) $ defined for $k\in {\mathbb
R}^d, w\in {\mathbb C}$ such that $\mathfrak{Re}\,w<\kappa
-\varepsilon \left( q\right) $. One can prove by induction, using
the formula (\ref{eq. 2.9b}), that, in every term $\Gamma ^k\mu _0$
of the expansion (\ref{4.55b}), the function $b_{n,p}^{(k)}\left(
z;q_1,...,q_{n-1};q\right) $ has, for fixed $q_1,...,q_{n-1}$, a
form similar to (\ref{4.55c}), i.e.
\begin{equation}\label{4.55d}
b_{n,p}^{(k)}\left( z;q_1,...,q_{n-1};q\right) = \varphi
^{(k)}_{q_1,...,q_{n-1}}
  \left( p-q;z-\varepsilon \left( q\right) \right) .
\end{equation}
Then, it follows that the coefficient functions $b_{n}\equiv
b_{n,p}$ of the solution $\mu $ given by (\ref{4.55b}) are of the
same form (\ref{4.55d}), in particular,
\begin{equation}\label{4.55e}
b_{2,p}\left( q_1;q\right) =\varphi _{q_1}
  \left( p-q;z-\varepsilon \left( q\right) \right)
\end{equation}
Plugging this expression into (\ref{eq. 2.21}), we find that $m_p$ depends only on the
differences $p-q$ and $z-\varepsilon \left( q\right)$:
\begin{equation}\label{4.45f}
 m_p\left( z,q\right) =\tau  \left( p-q;z-\varepsilon \left( q\right) \right) .
\end{equation}
Hence, by virtue of (\ref{eq. 1.10}), ( \ref{a-m-relation}) and
(\ref{4.45f}) the equation (\ref{eq. 4.5})
\begin{equation}\label{4.45h}
\frac{1}{2}{\left( p-q\right) }^2+ \varepsilon \left( q\right) + m_p\left(
\xi,q\right) =\xi
\end{equation}
writes as
\begin{equation}\label{4.45g}
\frac{1}{2}\left( p-q\right) ^2+\tau  \left( p-q;\gamma \right) =\gamma
\end{equation}
This implies that $\gamma \equiv \xi - \varepsilon \left( q\right)$ is a function of
$p-q$ alone.\hfill $\square $


Clearly, by the convexity of $e_{1,p}^0$ and the asymptotical properties of $%
m_p(\xi ,q)$ given in Corollary \ref{C3.4}, $%
G_p^{(1),\kappa }$ is a bounded domain, nonvoid for $\kappa
>\lambda _1(p)$,
and $\min \limits_{q\in G_p^{(1),\kappa }}\xi \left( q\right) =\lambda _1(p)$%
.

By the smoothness of $a_p\left( \xi ,q\right) $ with respect to both arguments, the
function $\xi \left( q\right) $ defined on $G_p^{\left( 1\right) ,\kappa }$ is also
smooth. Moreover, for $\alpha $ sufficiently small, this function has a unique
critical point (namely, a minimum), which is nondegenerate. In particular, it follows
that on every level of $\xi
\left( q\right) $,%
$$
\chi _x=\left\{ q\in G_p^{\left( 1\right) ,\kappa }:\;\xi \left( q\right)
=x\right\} ,
$$
one can define a measure $\nu _x$ (the Gelfand-Leray measure, see \cite
{Gelfand}), such that, for any integrable function $\varphi $ on $%
G_p^{\left( 1\right) ,\kappa }$,
\begin{equation}
\label{eq. *}\int_{G_p^{\left( 1\right) ,\kappa }}\varphi \left( q\right)
dq=\int\limits_{\lambda _1\left( p\right) }^\kappa dx\int_{\chi _x}\varphi
_xd\nu _x,
\end{equation}
where $\varphi _x=\varphi \mid _{\chi _x}$ is the restriction of
$\varphi $ to the surface $\chi _x$. From (\ref{eq. *}) it follows
that $L^2\left( G_p^{\left( 1\right) ,\kappa },dq\right) $ can be
represented as a direct integral of Hilbert spaces:
\begin{equation}
\label{eq. dir-int}L^2\left( G_p^{\left( 1\right) ,\kappa },dq\right)
=\int\limits_{\left[ \lambda _1(p),\kappa \right] }^{\oplus }{\mathcal{H}}_xdx,
\end{equation}
with ${\mathcal{H}}_x:=L^2\left( \chi _x,\nu _x\right) $.

Let us now consider the family $\left\{ F_{\xi (q)}^q\right\}_{q \in
G_p^{(1),\kappa }} \subset \left( {\cal B}^{(k)}\right) ^{\prime }$
of generalized eigenvectors of $H_p$. The next lemma , which may be
stated formally as an approximate orthonormality of this
family, is an important element of our constructions. We denote
$$
F\left( \varphi \right):= \int\limits_{G_p^\kappa }F_{\xi
(q)}^q\varphi (q)dq\in \left( {\cal B}^{(k)}\right) ^{\prime } ,
$$
for $\varphi \in {\cal D}\left( G_p^{(1),\kappa }\right)$, the
space of infinitely differentable functions with support
in $G_p^{(1),\kappa }$.
\begin{lemma}\label{L4.4}
(i) For any $\varphi \in {\cal D}\left( G_p^{(1),\kappa }\right) $,
one has $F\left( \varphi \right) \in {\cal F}$ .\\
(ii) There exist functions $S(q)$ \ and $M(q,q^{\prime })$ \ defined
for $q,q^{\prime }\in G_p^{(1),\kappa }$ , such that, for any
$\varphi _1,\varphi _2\in {\cal D}\left( G_p^{(1),\kappa }\right)$,
the following representation holds:
\begin{equation}
\label{eq. 4.21}
\begin{array}[b]{r}
\left( F\left( \varphi _1\right) ,F\left( \varphi _2\right) \right) _{
{\cal F}}=\int\limits_{\lambda _1\left( p\right) }^\kappa dx\left[
\int_{\chi _x}\left( 1+S_x(q)\right) \left| \varphi _x(q)\right| ^2d\nu
_x\right. \\ \left. +\int_{\chi _x}\int_{\chi _x}\bar \varphi
_x(q)M_x(q,q^{\prime })\varphi _x(q^{\prime })d\nu _x\left( q\right) d\nu
_x\left( q^{\prime }\right) \right] .
\end{array}
\end{equation}
Here, $\varphi _x,S_x$ \ and $M_x$ \ denote the
restrictions of the functions $\varphi ,S$ \ and $M$ \ to $\chi _x$%
 \ and $\chi _x\times \chi _x$ , respectively.\\
(iii) The following estimates hold with suitable constants $%
\bar C$, $\hat C$ :
\begin{equation}
\label{eq. 4.33'}\left| S(q)\right| \leq \bar C \, \frac \alpha
{\lambda _2(p)-\kappa },
\end{equation}
\begin{equation}
\label{eq. 4.33}\left| M(q,q^{\prime })\right| \leq \hat C
\,\,\alpha \, h(q)h(q^{\prime }),
\end{equation}
implying that $F\left( \varphi \right) \in {\cal F}$, for $%
\varphi \in L^2\left( G_p^{(1),\kappa }\right) $ \ and
\begin{equation}
\label{eq. 4.34}C_1\left\| \varphi \right\| _{L^2\left(
G_p^{(1),\kappa }\right) }\leq \left\| F(\varphi )\right\| \leq
C_2\left\| \varphi \right\| _{L^2\left( G_p^{(1),\kappa }\right) }.
\end{equation}
\end{lemma}
\textit{Proof}: (i) This assertion will follow from the
calculations
below.\\
(ii) In the sense of distributions, equation (\ref{eq.
4.21}) means that
\begin{equation}
\label{eq. 4.21'}\left( F_{\xi (q)}^q,F_{\xi (q^{\prime })}^{q^{\prime
}}\right) _{{\cal F}}=\left( 1+S(q)\right) \delta \left( q-q^{\prime
}\right) +M(q,q^{\prime })\delta \left( \xi (q)-\xi (q^{\prime })\right)
\end{equation}
Before we proceed, the following remarks are in order:\\
$1^{\circ}$. We seemingly make an abuse in calculating the scalar
product $\left( F_\xi ^q,F_{\xi ^{\prime }}^{q^{\prime }}\right)
_{{\cal F}}$ of two generalized functions. Such calculations can be
justified in the following way. The Fourier transform $\tilde F_{\xi
,n}^q\left( \zeta _1,...,\zeta _n\right) $ of the generalized
function $F_{\xi ,n}^q\left( q_1,...,q_n\right) $ is, as one can
easily verify, a usual function of the variables $\left( \zeta
_1,...,\zeta _n\right) $ polynomially bounded at infinity in these
variables. If, further, we view the scalar product $\left( F_{\xi
,n}^q,F_{\xi ^{\prime },n}^{q^{\prime }}\right) _{L^2\left( {\mathbb R}%
^{nd}\right) }$ as the limit of scalar products:%
$$
\left( \tilde F_{\xi ,n}^q,\tilde F_{\xi ^{\prime },n}^{q^{\prime }}\right)
_{L^2\left( {\mathbb R}^{nd}\right) }:=\lim \limits_{\delta \searrow 0}\int_{%
{\mathbb R}^{nd}}\overline{\tilde F_{\xi ,n}^q\left( \zeta _1,...,\zeta _n\right)
}\tilde F_{\xi ^{\prime },n}^{q^{\prime }}\left( \zeta _1,...,\zeta _n\right)
\prod\limits_{i=1}^n\left( e^{-\delta \left| \zeta _i\right| }d\zeta _i\right) , $$
then one can prove that this limit exists in the sense of convergence of generalized
functions of the variables $q,q^{\prime }$. We shall not provide the details of this
justifying procedure, and write instead directly its result entering our
calculations.\\
$2^{\circ}$. We shall exploit the ''orthogonality'' of the generalized eigenfunctions $%
F_{\xi (q)}^q$, $q\in G_p^{(1),\kappa }$, corresponding to different
eigenvalues $\xi (q)\neq \xi (q^{\prime })$, by supposing that the
support
of the generalized function%
$$
Q\left( q,q^{\prime }\right) :=\left( F_{\xi (q)}^q,F_{\xi (q^{\prime
})}^{q^{\prime }}\right) _{{\cal F}}
$$
is contained in the surface $\Sigma =\left\{ \left( q,q^{\prime }\right) \in
G_p^{(1),\kappa }\times G_p^{(1),\kappa }:\;\xi (q)=\xi (q^{\prime
})\right\} $:
\begin{equation}
\label{eq. **}\text{supp}Q\subset \Sigma
\end{equation}
and neglect in our calculation all terms which do not contribute to the
factors in front of $\delta \left( \xi (q)-\xi (q^{\prime })\right) $ or $%
\delta \left( q-q^{\prime }\right) $. However, the relation (\ref{eq. **})
likewise needs a justification. Namely, if we did not skip these
''non-contributing'' terms in the calculations, we would obtain a
generalized function $\tilde Q\left( q,q^{\prime }\right) $, such that, for
smooth functions $\varphi _i\left( q\right) $, $i=1,2$ with support
contained in $G_p^{(1),\kappa }$, the scalar product%
$$
\left( F\left( \varphi _1\right) ,F\left( \varphi _2\right) \right) _{{\cal F%
}}=\int\limits_{G_p^{(1),\kappa }\times G_p^{(1),\kappa }}\tilde Q\left(
q,q^{\prime }\right) \bar \varphi _1\left( q\right) \varphi _2\left(
q^{\prime }\right) dqdq^{\prime }
$$
would be finite, in particular, $F\left( \varphi \right) \in {\cal F}$ for
smooth $\varphi $. On the other hand, if the support of $\varphi $ was
contained in an $\varepsilon $-neighbourhood of the level $\chi _x$, then $%
F\left( \varphi \right) \in E\left( x-\varepsilon ,x+\varepsilon \right)
{\cal F}$, where $\left\{ E\left( \Delta \right) \right\} $ denotes the
family of spectral projections of $H_p$. Hence, for $\varphi _i\left(
q\right) $, $i=1,2$ with supports respectively contained in nonintersecting $%
\varepsilon $-neighbourhoods of the levels $\chi _{x_i}$, where $x_1\neq x_2$%
, the vectors $F\left( \varphi _i\right) $, $i=1,2,$ would be
orthogonal. This proves in fact (\ref{eq. **}).

Keeping these remarks in mind, we proceed with the proof of (ii).\\
One has
\begin{equation}
\label{eq. 4.22}\left( F_{\xi (q)}^q,F_{\xi (q^{\prime })}^{q^{\prime
}}\right) _{{\cal F}}=\left( \Pi _1F_{\xi (q)}^q,\Pi _1F_{\xi (q^{\prime
})}^{q^{\prime }}\right) _{{\cal H}^{\left( \leq 1\right)
}}+\sum\limits_{n=2}^\infty \left( F_{\xi (q),n}^q,F_{\xi (q^{\prime
}),n}^{q^{\prime }}\right) _{{\cal H}^{\left( n\right) }}
\end{equation}
As $H_p$ is self-adjoint and $F_{\xi (q)}^q$ are its generalized
eigenfunctions with eigenvalue $\xi (q)$, the support of this
distribution is contained in $\xi (q)=\xi (q^{\prime })$. As $\Pi
_1F_{\xi (q)}^q=\psi ^q$ and $\Pi _1F_{\xi (q^{\prime })}^{q^{\prime
}}=\psi ^{q^{\prime }}$ are generalized eigenvectors of the operator
$A\left( \xi \right) $ for $\xi =\xi \left( q\right) =\xi \left(
q^{\prime }\right) $, we can use the relation (\ref{eq. 3.28a}):
\begin{equation}
\label{eq. 4.23}\left( \Pi _1F_{\xi (q)}^q,\Pi _1F_{\xi (q^{\prime
})}^{q^{\prime }}\right) _{{\cal H}^{\left( \leq 1\right) }}=\delta \left(
q-q^{\prime }\right) ,\;\left( \xi (q)=\xi (q^{\prime })\right) .
\end{equation}
We are therefore left with calculating $\left( F_{\xi (q),n}^q,F_{\xi
(q^{\prime }),n}^{q^{\prime }}\right) _{{\cal H}^{\left( n\right) }}$ for $%
n\geq 2$. To this aim, use is made of the representation (\ref{eq. 2.8})
\begin{equation}
\label{eq. 4.24}
\begin{array}[b]{r}
F_{\xi (q),n}^q=\sum\limits_{i=1}^nb_n\left( \xi (q);q_1,...,\check
q_i..,q_n;q_i\right) f_{\xi (q),1}^q\left( q_i\right) \\
+\int d_n\left( \xi (q);q_1,...,q_n;q^{\prime }\right) f_{\xi (q),1}^q\left(
q^{\prime }\right) dq^{\prime }.
\end{array}
\end{equation}
The second term in (\ref{eq. 4.24}) is a smooth function of $q_1,...,q_n$
and does not contribute to the terms containing $\delta \left( \xi (q)-\xi
(q^{\prime })\right) $ or $\delta \left( q-q^{\prime }\right) $. Likewise,
it is not hard to see that the only contributions to such terms come from
\begin{equation}
\label{eq. 4.25}
\begin{array}{r}
\int
\overline{b_n\left( \xi (q);q_1...\check q_i...q_n;q_i\right) f_{\xi
(q),1}^q\left( q_i\right) }b_n\left( \xi (q^{\prime });q_1...\check
q_i...q_n;q_i\right) f_{\xi (q^{\prime }),1}^{q^{\prime }}\left( q_i\right)
dq_1...dq_n \\ =\int g_n\left( \xi (q),\xi (q^{\prime });\hat q\right)
\overline{f_{\xi (q),1}^q\left( \hat q\right) }f_{\xi (q^{\prime
}),1}^{q^{\prime }}\left( \hat q\right) d\hat q,
\end{array}
\end{equation}
where
\begin{equation}
\label{eq. 4.26}g_n\left( \xi ,\xi ^{\prime };\hat q\right) =\int \overline{%
b_n\left( \xi ;q_1...q_{n-1};\hat q\right) }b_n\left( \xi ^{\prime
};q_1...q_{n-1};\hat q\right) dq_1...dq_{n-1}.
\end{equation}

In the integral over $\hat q$ in the r.h.s. of equation (\ref{eq.
4.25}), we separate the singular parts of $f_{\xi (q),1}^q$, $f_{\xi
(q^{\prime }),1}^{q^{\prime }}$ using the \textit{Sokhotski
formula}:
\begin{equation*}
\frac{1}{x+i0} = \mathcal{P}\left( \frac{1}{x}\right) +i\pi \delta
\left( x\right)
\end{equation*}
in their expression (\ref {eq. 3.25}) and the fact that $a_p\left(
\xi \left( q\right) ,q^{\prime }\right) =\xi \left( q\right) $
implies $\xi \left( q\right) =\xi \left( q^{\prime }\right) $, hence
also $a_p\left( \xi \left( q\right) ,q^{\prime }\right) =\xi \left(
q^{\prime }\right) $ (in view of the uniqueness of the solution of
$a_p\left( \xi ,q\right) =\xi $):
\begin{equation}
\label{eq. 4.27}
\begin{array}{r}
f_{\xi (q),1}^q\left( q^{\prime }\right) =\delta \left( q-q^{\prime }\right)
+i\pi R_{\xi (q)}\left( q,q^{\prime }\right) \delta \left( \xi (q)-\xi
(q^{\prime })\right) \\
+\,{\rm regular\;terms}
\end{array}
\end{equation}
where
\begin{equation}
\label{eq. 4.28}
\begin{array}{l}
R_\xi \left( q,q^{\prime }\right) =\alpha ^2K_\xi ^{-}\left( \xi
;q,q^{\prime }\right) -\alpha f_{\xi ,0}^qc(p-q,q) \\
+\alpha ^2f_{\xi ,0}^q\int K_\xi ^{-}\left( \xi ;q^{\prime },q^{\prime
\prime }\right) c(p-q^{\prime \prime },q^{\prime \prime })\left( a_p\left(
\xi ,q^{\prime \prime }\right) -\xi +i0\right) ^{-1}dq^{\prime \prime }.
\end{array}
\end{equation}
The regular parts do not contribute to (\ref{eq. 4.25}), which becomes,
after performing the integration over $\hat q$:
\begin{equation}
\label{eq. 4.29}
\begin{array}[b]{r}
g_n\left( \xi (q),\xi (q);q\right) \delta \left( q-q^{\prime }\right)
\hfill \\ +i\pi \delta \left( \xi (q)-\xi (q^{\prime })\right) \left[
g_n\left( \xi (q),\xi (q);q^{\prime }\right) R_{\xi (q)}\left( q,q^{\prime
}\right) \right.
\hfill \\ \left. -g_n\left( \xi (q),\xi (q);q\right) \bar R_{\xi (q)}\left(
q^{\prime },q\right) \right] \\
+\pi ^2\int\limits_{\xi (q^{\prime \prime })=\xi (q)}g_n\left( \xi (q),\xi
(q);q^{\prime \prime }\right) R_{\xi (q)}\left( q,q^{\prime \prime }\right)
\bar R_{\xi (q)}\left( q^{\prime },q^{\prime \prime }\right) dq^{\prime
\prime }.
\end{array}
\end{equation}
Let us define the function:
\begin{equation}
\label{eq. 4.30}T\left( \xi ,q^{\prime }\right) =\sum\limits_{n=2}^\infty
\frac 1{n!}ng_n\left( \xi ,\xi ;q^{\prime }\right) .
\end{equation}
Then, one can see from equations (\ref{eq. 4.23}),
(\ref{eq. 4.29}) that (\ref {eq. 4.21'}) is fulfilled with
\begin{equation}
\label{eq. 4.31}S(q)=T(\xi (q),q),
\end{equation}
\begin{equation}
\label{eq. 4.32}
\begin{array}{c}
M(q,q^{\prime })=i\pi \left[ T(\xi (q),q^{\prime })R_{\xi (q)}\left(
q,q^{\prime }\right) -T(\xi (q),q)\bar R_{\xi (q)}\left( q^{\prime
},q\right) \right] \\
+\pi ^2\int T(\xi (q),q^{\prime \prime })R_{\xi (q)}\left( q,q^{\prime
\prime }\right) \bar R_{\xi (q)}\left( q^{\prime },q^{\prime \prime }\right)
dq^{\prime \prime }.
\end{array}
\end{equation}
(iii) Using the estimates in Lemma \ref{L3.2}, one obtains for $T$:%
$$
\left| T\left( \xi ,q^{\prime }\right) \right| \leq C \,\frac
{\alpha} {\lambda _2(p)-\kappa }\,\,\sum\limits_{n=2}^\infty
\,\,\frac{\left\| h\right\| _{L_2}^{2(n-1)}}{\left( n-1\right)
!}=\bar C \,\,\frac \alpha {\lambda _2(p)-\kappa } \,.
$$
Also, from the inequalities (\ref{eq. 4.17}) and (\ref{eq. 3.8}) it follows
that%
$$
\left| R_{\xi (q)}\left( q,q^{\prime }\right) \right| \leq Ch(q)h(q^{\prime
}),
$$
wherefrom (\ref{eq. 4.33'}), (\ref{eq. 4.33}) follow. Using these estimates
in equation (\ref{eq. 4.21}), one obtains that $F(\varphi )\in {\cal F}$ for any $%
\varphi \in L^2\left( G_p^{(1),\kappa }\right) $ and, moreover the estimate (%
\ref{eq. 4.34}) holds. The lemma is proved.       \hfill $\square$

Let now ${\cal H}_1^\kappa (p)\subset {\cal F}$ be the subspace
spanned by $ \left\{ F(\varphi ),\varphi \in L^2\left(
G_p^{(1),\kappa }\right) \right\} $. Equation (\ref{eq.
4.34}) implies that the application $\varphi \longmapsto
F\left( \varphi \right) $ is continuous and invertible. Thereby, ${\cal H}%
_1^\kappa (p)$ is $H_p$-invariant and
\begin{equation}
\label{eq. 4.35}H_pF\left( \varphi \right) =F\left( \hat \xi \varphi \right)
,
\end{equation}
where
\begin{equation}
\label{eq. 4.36}\left( \hat \xi \varphi \right) (q)=\xi (q)\varphi (q).
\end{equation}
\begin{lemma}\label{4.5}  There exists a bounded, invertible operator $
B:L^2\left( G_p^{(1),\kappa }\right) \rightarrow L^2\left(
G_p^{(1),\kappa }\right) $ \ which commutes with $H_p$ \ and such
that:
\begin{equation}
\label{eq. 4.37}\left( F\left( B\varphi _1\right) ,F\left( B\varphi
_2\right) \right) _{{\cal F}}=\left( \varphi _1,\varphi _2\right)
_{L^2\left( G_p^{(1),\kappa }\right) }.
\end{equation}
\end{lemma}
\textit{Proof}:  We use the representation (\ref{eq. dir-int}) of
$L^2\left( G_p^{(1),\kappa },dq\right) $ as a direct integral of the
spaces ${\mathcal{H}}_x$ and write ${\cal H}_1^\kappa (p)$ as a
direct integral:
\begin{equation}
\label{eq. 4.38}{\cal H}_1^\kappa (p)=\int\limits_{\left[ \lambda
_1\left( p\right) ,\kappa \right] }^{\oplus }{\cal H}_{1,x}dx \,.
\end{equation}
Here ${\cal H}_{1,x}$ is the image of ${\mathcal{H}}_x$ by the application of $L^2\left(
G_p^{(1),\kappa },dq\right) $ into ${\cal H}_1^\kappa (p)$ and consists of functionals
\begin{equation}
\label{eq. 4.39}F_x\left( \varphi \right) =\int\limits_{\chi _x}F_{\xi
\left( q\right) }^q\varphi \left( q\right) d\nu _x(q).
\end{equation}
By virtue of (\ref{eq. 4.21'}),
\begin{equation}
\label{eq. 4.40}
\begin{array}[b]{r}
\left( F_x\left( \varphi _1\right) ,F_x\left( \varphi _2\right) \right) _{
{\cal F}}=\int\limits_{\chi _x}\left( 1+S(q)\right) \bar \varphi
_1(q)\varphi _2(q)d\nu _x(q) \\ +\int\limits_{\chi _x\times \chi _x}\bar
\varphi _1(q)M(q,q^{\prime })\varphi _2(q^{\prime })d\nu _x(q)d\nu
_x(q^{\prime }) \\
=\left( \left( I_x+V_x\right) \varphi _1,\varphi _2\right) _{{\cal H}%
_{1,x}},
\end{array}
\end{equation}
where $I_x$ is the unit operator in ${\cal H}_{1,x}$ and
$V_x$ is a bounded operator with small norm (cf. equations (\ref{eq.
4.33'}), (\ref{eq. 4.33})). Also, $H_p$ acts in ${\cal
H}_{1,x}$ as $xI_x$.

Let $B_x=\left( I_x+V_x\right) ^{-1/2}$. Then, equation (\ref{eq.
4.40}) reads as
$$\left( F_x\left( B_x\varphi _1\right) ,F_x\left(
B_x\varphi _2\right) \right) _{{\cal F}}=\left( \varphi _1,\varphi
_2\right) _{{\cal H}_{1,x}}.$$ Finally, defining $B =
\int\nolimits_{\left[ \lambda _1(p),\kappa \right] }^{\oplus
}B_xdx$, one gets both that the operator $B$ commutes with $H_p$ and
that equation (\ref{eq. 4.37}) is satisfied . \hfill $\square$

\section{The ground state of $H_p$}

A detailed analysis of the ground state of $H_p$ is performed in
arbitrary dimension in \cite{Minlos}. In this section we shall
briefly show how the existence of the ground state follows from our
considerations for $d\geq 3$.

As explained in Section \ref{outline}, $H_p$ has a ground state if,
and only
if, there exists $\xi <\lambda _1\left( p\right) $, such that operator $%
A_p\left( \xi \right) $ in ${\cal H}^{\left( \leq 1\right) }$ has the eigenvalue $\xi $.
By the analysis done in Section \ref{sec:fried}, $A_p\left( \xi \right) $ has
one simple eigenvalue $e_p\left( \xi \right) <\lambda _1\left( p\right) $ if, and only
if, $\Delta _p\left( \lambda _1\left( p\right) \right) <0$ (where
$\Delta_p\left( \xi \right) $ is the function defined by equation
(\ref {eq. 3.21}) for
$A_p\left( \xi \right) $), in which case $e_p\left( \xi \right) $ equals the unique
solution of the equation $\Delta _p\left( \xi \right) =0$. Since $e_p^{\left(
0\right) }-\lambda _1\left( p\right) \rightarrow \infty $ for $p\rightarrow \infty $,
while $\left( v,R_B\left( \lambda _1\left( p\right) \right) v\right) $ (with $v$ and $B$
corresponding to $A_p\left( \xi \right) $) is bounded, $\left\{ p;\Delta _p\left(
\lambda _1\left( p\right) \right) <0\right\} $ is a bounded domain.

As seen from equation (\ref{eq. 1.25}), $A_p\left( \xi \right) $ is
a decreasing
family (in the usual order of self-adjoint operators), implying that $%
e_p\left( \xi \right) $ is a decreasing function of $\xi \in \left( -\infty
,\lambda _1\left( p\right) \right) $. We conclude that the equation $%
e_p\left( \xi \right) =\xi $ has a solution $\xi _p^{\left( 0\right)
}$ if, and only if, $p$ belongs to the subdomain
$$
G^{\left( 0\right) }=\left\{ p:\;e_p\left( \lambda _1\left( p\right) \right)
<\lambda _1\left( p\right) \right\} .
$$
For $p\in G^{\left( 0\right) }$, let $F_{\xi _p^{\left( 0\right)
},1}$ be an
eigenvector of $A_p\left( \xi _p^{\left( 0\right) }\right) $, and $%
F_{\xi _p,2}\in {\cal H}^{\left( \geq 2\right) }$ be defined according to (\ref
{eq. 1.50}). Then, the vector $F_p^{\left( 0\right)
}=F_{\xi _p^{\left( 0\right) },1} + F_{\xi _p^{\left(
0\right) },2}$ is a ground state of the operator $H_p$.

Therefore, $H_p$ has a unique ground state if $p\in G^{\left(
0\right) }$, and no ground state if $p\notin G^{\left( 0\right) }$.
\section{Concluding remarks}

The main result of the paper is the construction, in the weak
coupling regime, of a manifold of states indexed by a phonon
momentum $q$. The ground state describing a single polaron becomes
unstable at a certain momentum threshold, above which it dissolves
into this manifold. It is to be expected that at still higher
momenta the latter states become themselves unstable and dissolve
into two-phonon states, etc. The representation (\ref{eq. 4.55a}) of
the eigenvalue $\xi \left( q\right) $ strongly suggests the
interpretation of the generalized eigenfunctions $F_{\xi \left(
q\right) }^q$ associated to it as scattering states of a free phonon
and a certain particle with the dispersion law $\gamma \left(
k\right) $. We cannot yet decide whether the latter particle is a
polaron defined in our Theorem \ref{T1.1}, i.e. whether $\gamma
\left( k\right) =\xi _k^{\left( 0\right) } $, although we checked
that this equality is true in the first nontrivial order in coupling
constant: $\sim \alpha^2$. In this case the ground state instability
at high $k$ might be interpreted as emission of a phonon.

Unfortunately, we were not able to prove in the present paper two essential results in
favour of the above heuristic physical picture:

1. First of all, we did not construct the whole
one-boson subspace ${\cal H}_1^{\kappa =\lambda _2\left( p\right) }$
up to the two-boson threshold $\lambda _2\left( p\right) $. The
approach used here of eliminating the higher components of
the eigenvectors can equally well be applied in the case of the
decomposition ${\cal F}={\cal H}^{\leq 2}\oplus {\cal H}^{\geq 3}$,
leading to a family of self-adjoint operators $\left\{ A\left( \xi
\right) ,\;\lambda _1\left( p\right) \leq \xi \leq \kappa \right\}
$, (where $\lambda _2\left( p\right) \leq \kappa <\lambda
_3\left( p\right) $, i.e. $\kappa $ is between the two-boson and the
three-boson threshold, defined similarly with $\lambda _2\left(
p\right) $), acting in the space ${\cal H}^{\leq 2}$ of triples
$\left( f_0,f_1\left( \cdot \right) ,f_2\left( \cdot ,\cdot \right)
\right) $. These operators have a more complicated structure than
the Friedrichs operators in ${\cal H}^{\leq 1}$ and their spectral
analysis and scattering theory is not available in such
details as for the Friedrichs operators. If this theory was
elaborated (e.g. using equations analogous to the
Faddeev-Yakubovski equations for the resolvent of $n$-body
Schr\"odinger operators, see \cite{Yakubovski} - \cite{Hepp}), then
the approach of the present paper would provide the
construction of the whole one-boson subspace and of a part of the
two-boson subspace.

2. Secondly, we did not prove the
completeness of the constructed subspaces ${\cal H}%
_0\left( p\right) $ (generated by the ground state) and ${\cal H}_1^\kappa
\left( p\right) $, meaning that in $\left( {\cal H}_0\left( p\right) \oplus
{\cal H}_1^\kappa \left( p\right) \right) ^{\perp }$ the spectrum of $H_p$
has no point below $\kappa $. We are convinced that this assertion is true
and hope to prove it in the future.

\vspace {0.5cm}

\noindent {\bf Acknowledgements}

\vspace {0.1cm}

\noindent N.A. and R.A.M. acknowledge the warm hospitality of the C.P.T.
Luminy-Marseille, where the project of this work was born. R.A.M. also acknowledges
financial support from the Scientific Fund of the Russian Federation (Grant No.
02-01-00444) and the Presidential Fund for Support of Scientific Schools of Russia
(Grant No. 934 2003.1).

\end{document}